\documentclass[useAMS,usenatbib]{mn2e}

\usepackage{graphicx,bm,amsmath,multirow,subfigure,longtable,float,color}

\usepackage[T1]{fontenc}
\usepackage{lmodern}
\usepackage{amsfonts}

\newcommand{\Msol}{\mbox{$M_{\sun}$}}

\newcommand{\Cov}{\mbox{$\mathrm{Cov}$}}

\newcommand{\Mpc}{\mbox{$\mathrm{Mpc}$}}
\newcommand{\Mpch}{\mbox{$h^{-1}\mathrm{Mpc}$}}

\newcommand{\tr}{\mbox{$\mathrm{tr}$}}

\newcommand{\erfc}{\mbox{$\mathrm{erfc}$}}

\title[Cosmic variance of cluster weak lensing]{Cosmic variance of the galaxy cluster weak lensing signal}
\author[D. Gruen et al.]
{\parbox{\textwidth}{D. Gruen$^{1,2}$\thanks{E-mail:
dgruen@usm.uni-muenchen.de (DG)}, S. Seitz$^{1,2}$, M. R. Becker$^{3,4}$, O. Friedrich$^{1,2}$ and A. Mana$^{1,2}$}\vspace{0.4cm}\\
\parbox{\textwidth}{$^{1}$University Observatory Munich, Scheinerstrasse 1, 81679 Munich, Germany\\
$^{2}$Max Planck Institute for Extraterrestrial Physics, Giessenbachstrasse, 85748 Garching, Germany\\
$^{3}$KIPAC, Physics Department, Stanford University, Stanford, CA 94305, USA\\
$^{4}$KIPAC, SLAC National Accelerator Laboratory, Menlo Park, CA 94025, USA\\
}}

\begin{document}

\date{}

\pagerange{\pageref{firstpage}--\pageref{lastpage}} \pubyear{2014}

\maketitle

\label{firstpage}

\begin{abstract}
Intrinsic variations of the projected density profiles of clusters of galaxies at fixed mass are a source of uncertainty for cluster weak lensing. We present a semi-analytical model to account for this effect, based on a combination of variations in halo concentration, ellipticity and orientation, and the presence of correlated haloes. We calibrate the parameters of our model at the 10 per cent level to match the empirical cosmic variance of cluster profiles at $M_{200m}\approx10^{14}\ldots10^{15}h^{-1}\Msol$, $z=0.25\ldots0.5$ in a cosmological simulation. We show that weak lensing measurements of clusters significantly underestimate mass uncertainties if intrinsic profile variations are ignored, and that our model can be used to provide correct mass likelihoods. Effects on the achievable accuracy of weak lensing cluster mass measurements are particularly strong for the most massive clusters and deep observations (with $\approx20$ per cent uncertainty from cosmic variance alone at $M_{200m}\approx10^{15}h^{-1}\Msol$ and $z=0.25$), but significant also under typical ground-based conditions. We show that neglecting intrinsic profile variations leads to biases in the mass-observable relation constrained with weak lensing, both for intrinsic scatter and overall scale (the latter at the 15 per cent level). These biases are in excess of the statistical errors of upcoming surveys and can be avoided if the cosmic variance of cluster profiles is accounted for.
\end{abstract}

\begin{keywords}
gravitational lensing: weak -- galaxies: clusters: general -- cosmology: observations
\end{keywords}

\section{Introduction}

The largest objects ever formed depend most sensitively on small changes in the overall properties of the Universe. It is for this reason that clusters of galaxies are a versatile probe of cosmology. The number density of clusters and its evolution with redshift is influenced by the expansion history, the density and level of inhomogeneity of matter in the Universe and the growth of structures by means of gravitation. Both the parameters of a standard $\Lambda$CDM model and deviations from primordial Gaussianity or General Relativity are therefore accessible to cluster cosmology (cf., e.g., \citealt{2004MNRAS.353..457A,2009ApJ...692.1060V,2010ApJ...708..645R,2010MNRAS.406.1759M,2013MNRAS.434..684M,2013ApJ...763..147B,p2013cosmology,2015MNRAS.446.2205M} for individual analyses and \citealt{2011ARA&A..49..409A} for a recent review).

The building blocks of such analyses are (i) a cluster catalog with well-defined selection function based on, e.g. optical, X-ray or \citet[][hereafter SZ]{1972CoASP...4..173S} properties, (ii) a prediction, based on theory or simulations, of how the number density of clusters at given mass and redshift depends on the cosmological parameters in question \citep[e.g.][]{1974ApJ...187..425P,1999MNRAS.308..119S,2008ApJ...688..709T} and (iii) a mass-observable relation (MOR) that connects the observable from (i) to the mass-based prediction from (ii) in terms of a likelihood. The latter must not only describe the mean relation of mass and observable but, because of the observable limited nature of any practical survey, also the intrinsic scatter in the observable at fixed mass \citep[e.g.][]{2005PhRvD..72d3006L}. 

With the advent of large cluster catalogs, our imperfect knowledge of the MOR remains the most important limiting factor of cluster cosmology. It is difficult to predict or simulate, to the level of accuracy required, all physical effects that influence the observables, particularly ones that are of baryonic nature. Thus, one needs to rely on an empirical calibration of the MOR.

The latter can in principle be done by means of self-calibration, i.e. by constraining both cosmology and MOR parameters from the cluster catalog alone \citep{2003PhRvD..67h1304H,2004ApJ...613...41M,2005PhRvD..72d3006L}. However, this approach greatly reduces the cosmological power of cluster studies, especially when extensions to the most simple MOR and cosmological models are considered.

It is for this reason that weak lensing studies of clusters of galaxies are a particularly powerful complement to cluster cosmology. Being sensitive to all matter independent of its astrophysical state, lensing allows, in principle, an unbiased measurement of cluster mass. Several studies have used this approach to constrain MORs with lensing mass measurements of sets of individual clusters \citep[e.g.][]{2009ApJ...701L.114M,2012ApJ...754..119M,2012MNRAS.427.1298H,2012arXiv1208.0597V,2014MNRAS.442.1507G,2015MNRAS.446.2205M}, with great prospects of further improving statistics with on-going and future large surveys.

However, for accurate constraints on the mean MOR and its intrinsic scatter to be derived, one needs to ensure that the mass likelihood from a lensing analysis includes all actual sources of uncertainty. For typical cluster lensing studies, the dominant uncertainty is observational, i.e. based on the limited number of intrinsically elliptical background galaxies. Since all structures between sources and observer cause a lensing signal, an additional, irreducible uncertainty results from the variance of the matter density along the line of sight \citep[LOS; cf. e.g.][]{2001A&A...370..743H,2003MNRAS.339.1155H}. Most cluster lensing studies include either only the first or both of these effects when fitting the signature of a mass dependent density profile to observed data.

What this neglects, either entirely or apart from a low-dimensional approximation in terms of e.g. halo concentration, is the complexity of cluster profiles. The projected density profiles of two clusters at the same virial mass can differ greatly, as could be described accurately only with detailed information about the shape and orientation of their central dark matter halo and the positions and masses of subhaloes and neighbouring structures. Due to its relatively low resolution, no weak lensing analysis is able to uncover all or even most of this information. However, the uncertainty of mass measurement will depend on the statistical properties of these variations.

The goal of this work is to provide a model for the variations in projected density profiles of cluster of galaxies at fixed mass. We construct our model using analytical templates for the expected scatter due to variations in halo concentration, ellipticity and orientation, and its substructure and cosmic neighbourhood. We re-scale these templates to match the empirical variations of cluster profiles in a cosmological simulation, where both well-defined true masses and the lensing signal without any observational noise are known. We then use this semi-analytical covariance model to test and predict the accuracy of lensing measurements of cluster mass and MORs, and to assess the effect of neglecting intrinsic variations.

The structure of this paper is as follows. In Section~\ref{sec:sim} we briefly describe the simulations used. Section~\ref{sec:model} defines the components of our model for the mean cluster profile and, in particular, its intrinsic covariance. In Section~\ref{sec:covfit} we explain how the model is fitted to the simulated data. Effects of the intrinsic covariance on weak lensing cluster mass measurements and MOR studies are shown in Section~\ref{sec:effect}, before we summarize in Section~\ref{sec:summary}.

\subsection*{Conventions}

Our calculations are presented in a flat WMAP 7 year cosmology \citep{2011ApJS..192...18K} with $(\Omega_m,\Omega_b,\sigma_8,h,n)=(0.27,0.044,0.79,0.7,0.95)$, in consistency with the simulations used. The covariances can be readily re-scaled to different sets of cosmological parameters, up to a potential cosmology dependence of our model parameters. We explicitly write appropriate factors of $h=H_0/(100$~km s$^{-1}$ Mpc$^{-1})$ where applicable. We denote the radii of spheres around the cluster centre with fixed overdensity as $r_{\Delta m}$, where $\Delta=200$ is the overdensity factor of the sphere with respect to the mean matter density $\rho_m$ at the cluster redshift. The mass inside these spheres is labelled and defined correspondingly as $M_{\Delta m}=\Delta\times\frac{4\pi}{3}r_{\Delta m}^3\rho_m$.

We use a linear matter power spectrum $P_{\rm lin}(k,z)=D^2(z)P_{\rm lin}(k,0)$ with the transfer function model including baryonic effects of \citet{1998ApJ...496..605E} and normalize to our value of $\sigma_8$. For the linear growth factor $D(z)$, normalized to $D(0)=1$, we use the expression of \citet{3696}. We define the corresponding linear matter two-point correlation as
\begin{equation}
\xi_{\rm lin}(r,z)=\frac{1}{2\pi^2}\int dk k^2 P_{\rm lin}(k,z) \frac{\sin(kr)}{kr} \; . 
\end{equation}

For the scaling of some of the cluster properties we will use the common definition of peak height 
\begin{equation}
\nu=\delta_c/\sigma(M_{200m},z) \; , 
\end{equation}
where $\delta_c=1.686$ and 
\begin{eqnarray}
\sigma^2(M_{200m},z)= D^2(z) \nonumber \\ \times \frac{1}{2\pi^2}\int dk \; k^2\; P_{\rm lin}(k,0)\left|w\left(k,\left(\frac{3M_{200m}}{4\pi\rho_{0,m}}\right)^{1/3}\right)\right|^2 ,
\end{eqnarray}
with the Fourier transform of the top-hat window function of radius $r$, $w(k,r)$.

\section{Simulations}

\label{sec:sim}

We use the simulation labeled L1000W in \citet{2008ApJ...688..709T}. It consists of $1024^3$ dark matter particles of mass $6.98\times 10^{10} h^{-1}\Msol$ in a box of comoving size 1 $h^{-1}$ Gpc, simulated with the parallelized Adaptive Refinement Tree algorithm \citep{1997ApJS..111...73K,2008arXiv0803.4343G} from redshift $z=60$ to $z=0$, at an effective spatial resolution of $30 h^{-1}$~kpc. In a snapshot at redshift $z=0.24533$ almost 15,000 haloes at $0.95\times10^{14}h^{-1}\Msol\leq M_{200m}\leq1.5\times10^{15} h^{-1} \Msol$ are identified. 

The same haloes are also used in \citet{2010arXiv1011.1681B} and we employ the lensing maps computed in that work. For the selected haloes, the mass is integrated along the LOS and the lensing signal is determined on a grid of approximately $40$ comoving kpc$/$pixel using the Born approximation as described in \citet[Section~3]{2010arXiv1011.1681B}. All matter within comoving $\pm200\Mpch$ along the LOS and, transversely, in a square with comoving side length $20\Mpch$ centered on the cluster is included in the calculation. Background sources are assumed to be at a constant redshift $z_s=1$, where $\Sigma_{\rm crit}=4.22\times10^{15} h \Msol \Mpc^{-2}$. The density map inside a clustrocentric radius $\theta\leq1'$ is subject to resolution effects, which is why we discard it from our analysis.

\section{Model definition}

\label{sec:model}

In the following, we introduce our model for the covariance of projected cluster profiles. Its components, namely the model for the mean profile (Section~\ref{sec:meanprofile}) and the contributions to the covariance (Section~\ref{sec:compcov}), are described subsequently.

Consider the convergence profile of a given cluster with mass $M_{200m}=:M$ as
\begin{equation}
\bm{K}=\bm{\kappa}(M)+\bm{E} \; ,
\end{equation}
where $\bm{K}$ is the observed convergence profile and $\bm{\kappa}(M)$ is the mean convergence profile of clusters of fixed mass $M$. $\bm{E}$ is the residual between the two, which arises due to observational uncertainty and variations of the noiseless observable $\kappa$ profile at fixed mass. These variations could be due to uncorrelated structures along the LOS or the projected cluster density profile itself. The latter \emph{intrinsic} variations are the focus of this work.

The profiles are vectors measured in a system of radial bins, where bin $1\leq i\leq n$ is defined as $\theta\in[\theta_{i,\mathrm{min}},\theta_{i,\mathrm{max}}]$ without gaps, i.e. $\theta_{i,\mathrm{max}}=\theta_{i+1,\mathrm{min}}$.

Approximating the residuals $\bm{E}$ as a multivariate Gaussian with zero mean, one can write the likelihood for an observed $\bm{K}$ given a mass $M$ as 
\begin{eqnarray}
\mathcal{L}(\bm{K}|M)&=&\frac{1}{\sqrt{(2\pi)^n\det(C(M))}} \nonumber \\ &\times& \exp\left(-\frac{1}{2}\bm{E}^{\rm T}C^{-1}(M)\bm{E} \right) \; ,
\label{eqn:P}
\end{eqnarray}
where we have introduced the covariance $C$ of residuals $\bm{E}$. This $n\times n$ matrix is defined as
\begin{equation}
C_{ij}=\Cov(E_i,E_j) \; . 
\end{equation}

We make a semi-analytic ansatz for a parametric model of $C$ as
\begin{eqnarray}
C(M)&=&C^{\rm obs}+C^{\rm LSS}+c^{\rm conc}(\nu)C^{\rm conc}(M) \nonumber \\ 
&+&c^{\rm corr}(\nu)C^{\rm corr}(M)+c^{\rm ell}(\nu)C^{\rm ell}(M) \nonumber \\
\label{eqn:Cmodel}
\end{eqnarray}
with contributions from observational uncertainty (i.e., shape noise in shear measurement and Poisson noise in magnification studies) $C^{\rm obs}$, uncorrelated large scale structure along the LOS $C^{\rm LSS}$, scatter in halo concentration $C^{\rm conc}$, correlated secondary haloes near the cluster $C^{\rm corr}$ and variations in halo ellipticity and orientation $C^{\rm ell}$. The terms $C^{\rm obs}$ and $C^{\rm LSS}$ are independent of the cluster itself. In contrast to the common case where only these are considered, however, $C(M)$ inherits a mass dependence from the latter terms, which we will call \emph{intrinsic} variations of the cluster profile. They are described in detail in  Section~\ref{sec:compcov}. The empirical re-scaling factors $c^{\star}(\nu)$ of $C^{\rm corr}$, $C^{\rm ell}$ and $C^{\rm conc}$ are determined by fitting the covariance model to the simulated haloes (see Section~\ref{sec:covfit}).

\subsection{Mean profile}
\label{sec:meanprofile}

Several studies in the past have proposed functional forms for the three-dimensional density of dark matter haloes. \citet*[][hereafter NFW]{1997ApJ...490..493N} found that haloes are well-fit by the two-parametric broken power-law profile
\begin{equation}
\rho_{\rm NFW}(r)=\frac{\rho_0}{(r/r_{\rm s})(1+r/r_{\rm s})^2} \; ,
\label{eqn:nfw}
\end{equation}
with scale density $\rho_0$ and a scale radius $r_{\rm s}$ that can be expressed as a fraction of $r_{200m}$ using the concentration $c_{200m}=r_{200m}/r_{\rm s}$.

Two additional effects need to be accounted for in a realistic description of halo density profiles. Firstly, the enclosed mass of an NFW profile diverges logarithmically as $r\rightarrow\infty$. The density must therefore be truncated at large radii, for which a number of approaches have been proposed (e.g. \citealt{2003MNRAS.340..580T}; \citealt{2008MNRAS.388....2H}; \citealt*{2009JCAP...01..015B}, hereafter BMO). Secondly, the halo is embedded in an overdensity of correlated matter that contributes to the overall projected profile at all, and dominantly at large, radii. We closely follow the work of \citet{2011MNRAS.414.1851O}, who fit a superposition of a BMO profile with a linear two-halo term to simulated haloes, i.e.
\begin{equation}
\Sigma(M)=\kappa(M)\Sigma_{\rm crit}=\Sigma_{\rm 1h, BMO}(M)+\Sigma_{\rm 2h, lin}(M) \; .
\label{eqn:profilemodel}
\end{equation}
We describe the one-halo (Section~\ref{sec:onehalo}) and two-halo term (Section~\ref{sec:twohalo}) in the following.

\subsubsection{One-halo term}

\label{sec:onehalo}

BMO define a truncated version of the NFW profile as
\begin{equation}
\rho_{\rm BMO}(r)=\rho_{\rm NFW}(r)\times\left(\frac{\tau_{200m}^2}{\left(r/r_{200m}\right)^2+\tau_{200m}^2}\right)^{\beta} \; ,
\end{equation}
where we use $\beta=2$. Here we have introduced $\tau_{200m}$, the multiple of $r_{200m}$ around which the halo is smoothly truncated. The BMO profile can be analytically integrated to yield the projected profile $\Sigma_{\rm 1h, BMO}$, as well as the change in mass normalisation due to missing density relative to the NFW profile (cf. BMO; \citealt{2011MNRAS.414.1851O} and the \textsc{ample} code, \texttt{http://kipac.stanford.edu/collab/research/lensing/ ample/ample.c}).

Besides mass, our one-halo term has two additional free parameters, the concentration $c_{200m}$ and truncation radius $\tau_{200m}$. We fix the former with the mass-concentration relation of \citet{2008MNRAS.390L..64D}. \citet{2011MNRAS.414.1851O} note a weak dependence of $\tau_{\Delta_{\rm vir}m}$ on mass. Our scaling relation of $\tau_{200m}$ is determined by fitting the mean $\kappa$ profile of haloes in our simulation (cf. Section~\ref{sec:sim}) in logarithmic bins of $\Delta\log_{10}M_{200m}=0.05$ in the radial range of 3 to 15 arcmin using the full profile of eqn.~\ref{eqn:profilemodel}, with $\tau$ as a free parameter. The uncertainties for our $\chi^2$ minimization w.r.t. $\tau$ are derived from the ensemble variance of mean $\kappa$ in the respective mass bin. We find that the relation is well described as a linear function of $\nu$. Our fit in $\nu\in[2.2,3.7]$ yields
\begin{equation}
\tau_{200m}=\left\lbrace\begin{array}{cc}3.85-0.73\nu & \nu\leq3.7 \\ 1.15 & \nu>3.7 \end{array}\right. \; ,
\label{eqn:taum}
\end{equation}
where we have assumed a constant value for $\tau_{200m}$ above the peak height range covered by our simulated haloes.

We note that the fact that more massive haloes are, on average, truncated at smaller multiples of their $r_{200m}$ radii could be interpreted in terms of the steepening of the density slope with increasing mass accretion rate observed by \citet{2014ApJ...789....1D}.

\subsubsection{Two-halo term}

\label{sec:twohalo}

The two-halo surface mass density due to linear evolution around a cluster of mass $M$ at redshift $z$ can be written as (e.g. \citealt{2011MNRAS.414.1851O})
\begin{equation}
\Sigma_{\rm 2h, lin}(\theta)=b_{\rm h}(M, z)\,\rho_{c,0}\,\Omega_m \,W(\theta, z)\, D_A^{-2}(z) \; .
\label{eqn:2ht}
\end{equation}
Here, $b_{\rm h}(M, z)$ is the linear bias of the cluster halo (which we calculate according to \citealt{2010ApJ...724..878T}) and $D_A(z)$ is the angular diameter distance to redshift $z$. We have defined the projected linear excess Lagrangian depth $W(\theta, z)$ as
\begin{eqnarray}
W(\theta, z)&=& [D_A(z)(1+z)]^2 \nonumber \\ &\times& \int_{-\Delta\zeta}^{+\Delta\zeta} \xi_{\rm lin}\left(\sqrt{[D_A(z)(1+z)\theta]^2+\zeta^2},z\right)\mathrm{d}\zeta \nonumber \\
&=& \int\frac{l\mathrm{d}l}{2\pi}J_0(l\theta)P_{\rm lin}\left(\frac{l}{(1+z)D_A(z)},z\right) \; ,
\label{eqn:W}
\end{eqnarray}
where $\pm\Delta\zeta$ is the interval in comoving distance along the LOS integrated in our simulated projected density profiles. For large enough $\Delta\zeta$, almost all correlated matter is included in the integration and the latter equality holds. In eqn.~\ref{eqn:W}, we have used the Bessel function of the first kind and order zero, $J_0$. $W$ has units of comoving volume per solid angle and is readily interpreted as the excess in Lagrangian volume per solid angle that has moved to the projected vicinity of a structure with unit bias due to linear evolution.
 
The mass measurements of our haloes from the simulations contain all particles within $r_{200m}$. We therefore need to correct for contributions of two-halo matter $M_{\rm 2h}$ when defining the mass of the one-halo profile, which we do at first order as
\begin{equation}
M_{200m,1h}=M_{200m}-M_{\rm 2h}(r_{200m}, M_{200m}, z) \; . 
\end{equation}
Inside a sphere of radius $r$ around the cluster centre, the mass of two-halo matter is
\begin{equation}
M_{\rm 2h}(r, M, z)=b_{\rm h}(M, z)\,\rho_{c,0}\,\Omega_{m,0}\,U(r,z) \; .
\end{equation}
Here we have defined the linear excess Lagrangian volume inside a sphere of radius $r$ as
\begin{eqnarray}
U(r,z)&=&\int_0^{r(1+z)} \mathrm{d}r' \xi_{\rm lin}(r',z)\, 4\pi r'^2 \nonumber \\ &=& D^2(z)\, U(r(1+z),0) \; . 
\end{eqnarray}

\subsubsection{Comparison to simulations}

\begin{figure*}
\centering
\includegraphics[width=0.48\textwidth]{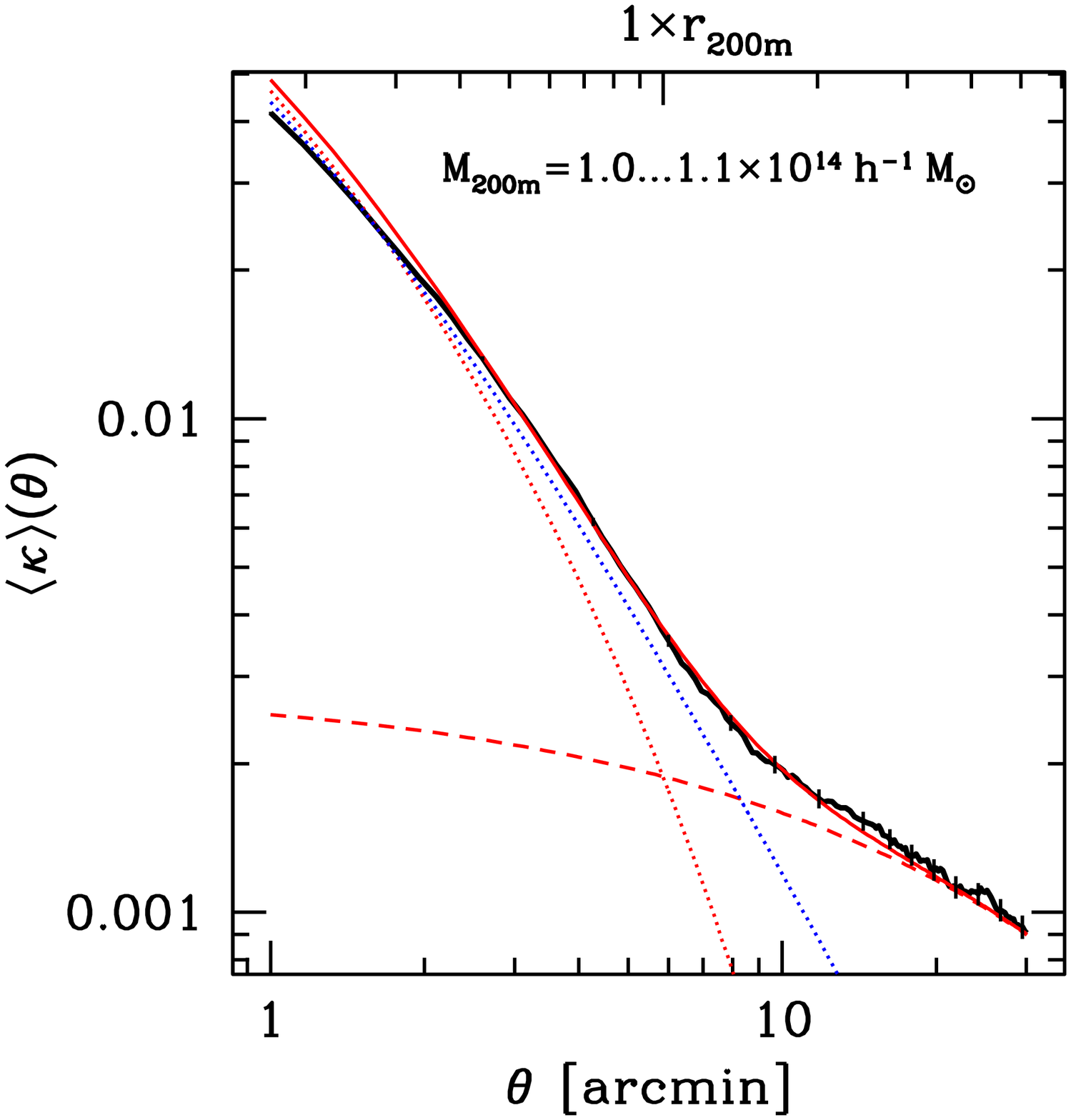}
\includegraphics[width=0.48\textwidth]{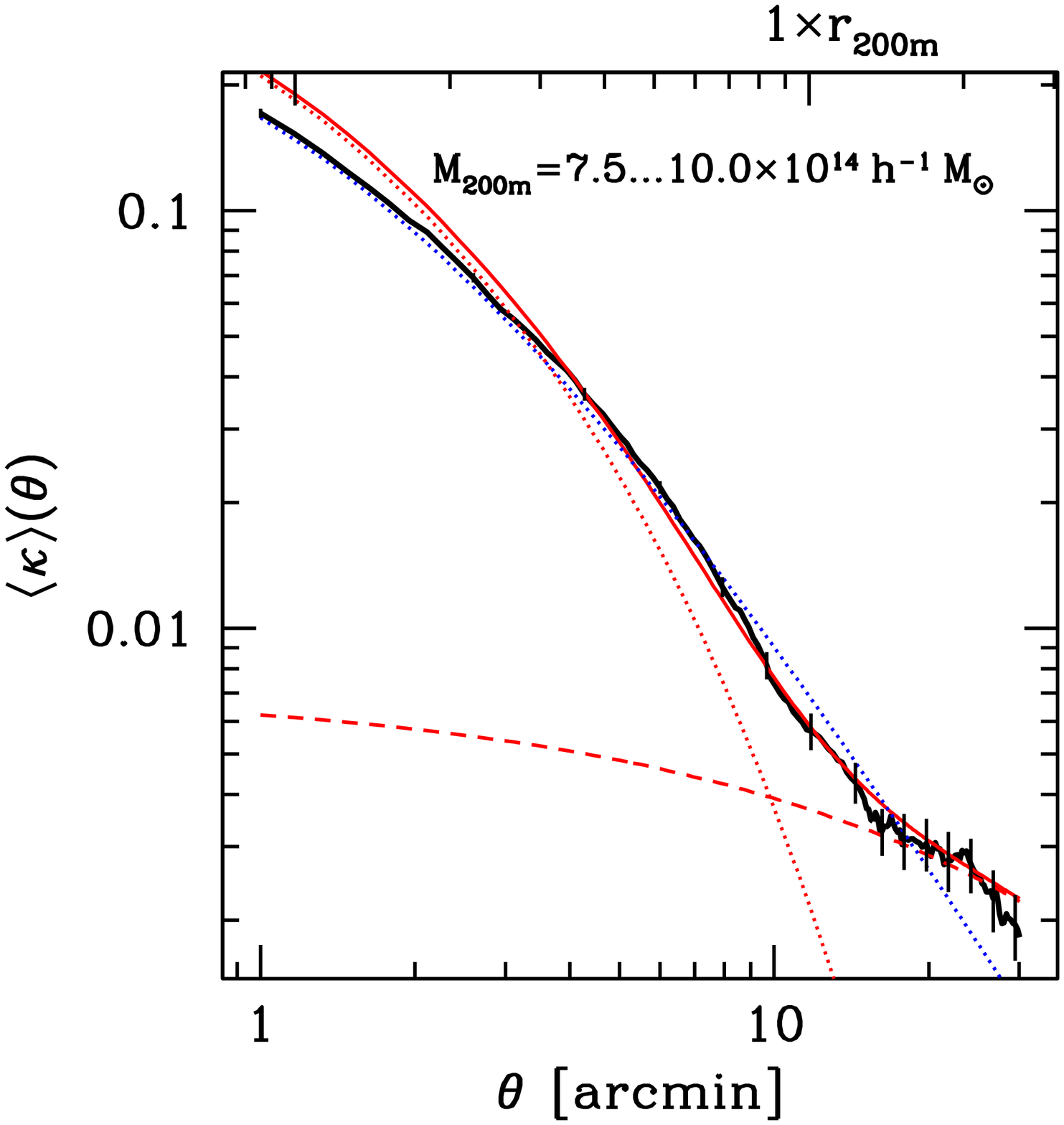}
\caption{Mean convergence profiles of haloes at the low-mass (left) and high-mass (right) end of our sample, shown for haloes at the simulation redshift $z=0.24533$ with corresponding angular (bottom axis) and $r_{200m}$ (top axis) scales. Black lines show average profiles in the simulations, with error bars giving the uncertainty of the mean. Our model is shown as the red, solid line, composed of the BMO one-halo term (red, dotted) with truncation radius according to equation~\ref{eqn:taum} and the two-halo term (red, dashed). The blue, dotted line shows un-truncated one-halo NFW profile for comparison.}
\label{fig:profile}
\end{figure*}

Figure~\ref{fig:profile} shows model and mean profiles of simulated haloes in two mass bins spanning most of the dynamic range of our halo catalog (see Section~\ref{sec:sim} for details on the simulations). The model fits the data well at projected $r>0.3\times r_{200m}$, but moderately overestimates projected density at smaller radii, where an un-truncated NFW profile without two-halo contribution is a better fit. Potential reasons for the discrepancy include factual deviations of the simulated dark matter haloes from the NFW profile at small radii (cf. \citealt{2010arXiv1011.1681B}, their Fig.~2, for an analysis based on the same cluster sample), resolution effects at small radii and the simplified nature of our linear superposition of the collapsed halo profile with a linear two-halo term (cf. \citealt{2008MNRAS.388....2H} for a different approach, in which the three-dimensional density is assumed to be piecewise equal to the NFW one-halo or a linear two-halo term, only). Since we only use the model for re-normalization of the ensemble mean profile to the mass of a given cluster (cf. Section~\ref{sec:ensemble}), this is not problematic for our purposes.

\subsection{Components of profile covariance}

\label{sec:compcov}

\begin{figure*}
\centering
\includegraphics[width=\textwidth]{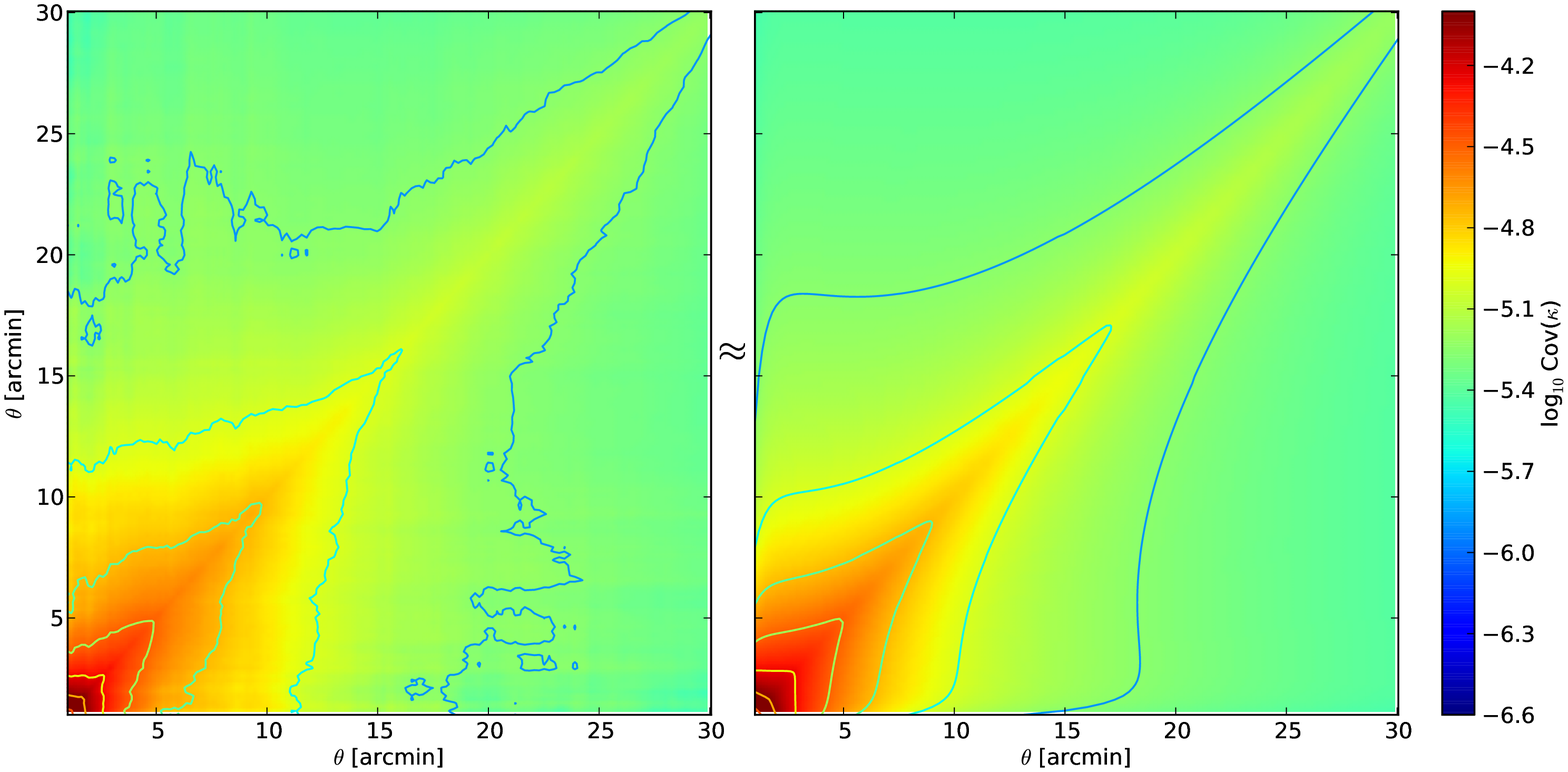}
\includegraphics[width=\textwidth]{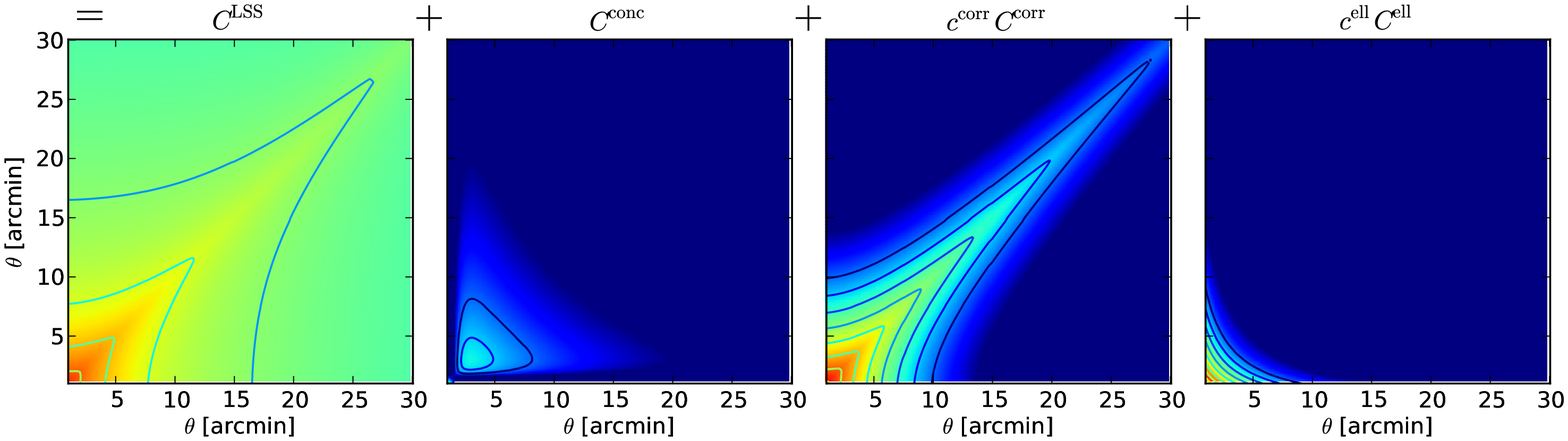}
\caption{Illustration of our model for the profile covariance. Top left panel shows the empirical intrinsic covariance of $\kappa$ profiles between 1 and 30~arcmin of simulated clusters of mass $M_{200m}=1.8\ldots2.2\times10^{14}\Msol$. Our model prediction for the covariance is shown in the top right panel. It is a linear combination of the components shown in the bottom panel. Even at this relatively low mass, uncorrelated large-scale structure (left in lower panel) is an insufficient description of the observed covariance, especially in the inner regions. We model the additional covariance as a linear combination of variation of the concentration parameter, the covariance of correlated haloes and halo ellipticity and orientation (remaining plots in lower panel). All panels use the same color scale with contour lines spaced by $\Delta\log_{10}\Cov\kappa=0.25$. At the mass plotted here, $c^{\rm corr}C^{\rm corr}$ is the dominant component of the intrinsic (co)variance on most scales, with significant contributions from $c^{\rm ell}C^{\rm ell}$ in the innermost region and subdominant concentration scatter. We note, however, that the relative importance of these three components changes as a function of mass (cf. Fig.~\ref{fig:var}).}
\label{fig:sketch}
\end{figure*}

We have made the ansatz for the residual between a given, noise-free cluster $\kappa$ profile $\bm{K}$ and the mean profile at its mass $\bm{\kappa}(M)$ as a sum of multivariate Gaussian vectors corresponding to four distinct physical effects: uncorrelated large scale structure along the LOS, variations in concentration, halo asphericity and orientation, and correlated haloes in the cluster substructure and neighbourhood (cf. eqn.~\ref{eqn:Cmodel}). All of these effects have the potential to bias the lensing mass measurement or, equivalently, can be interpreted as a source of intrinsic covariance between the components of $\bm{K}-\bm{\kappa}(M)$ that causes additional intrinsic noise on a cluster-by-cluster basis. In this section, we provide analytic expressions for all four components of the intrinsic covariance.

\subsubsection{Uncorrelated large-scale structure}

For a given source population, random structures along the LOS introduce a covariance in convergence measured in angular bins, which can be written as \citep[e.g.][]{1998MNRAS.296..873S,2003MNRAS.339.1155H,2011ApJ...738...41U}
\begin{equation}
C^{\rm LSS}_{ij}=\int \frac{l \mathrm{d}l}{2\pi} P_{\kappa}(l) \hat{J}_0(l\theta_i) \hat{J}_0(l\theta_j) \; .
\label{eqn:integral}
\end{equation}
Here, $\hat{J}_0(l\theta_i)$ is the area-weighted average of $J_0$ over annulus $i$. Using the identity (\citealt{Abramowitz}, their eqn.~9.1.30)
\begin{equation}
\left(\frac{1}{z}\frac{\rm d}{\mathrm{d}z}\right)^k [z^{l} J_{l}(z)] = z^{l-k}J_{l-k}(z) \; .
\end{equation}
with $k=l=1$ and integrating yields (cf. also \citealt{2011ApJ...738...41U}, their eqn.~16) 
\begin{equation}
\hat{J}_0(l\theta_i)=\frac{1}{2d l\theta_i}[(1+d)J_1(l\theta_i(1+d)) - (1-d)J_1(l\theta_i(1-d))] \; ,
\end{equation}
where we have assumed an annulus of width $2d\theta_i$ from $(1-d)\theta_i$ to $(1+d)\theta_i$.

The convergence power spectrum $P_{\kappa}$ in eqn.~\ref{eqn:integral} is obtained from the matter power spectrum by means of the \citet{1954ApJ...119..655L} approximation as
\begin{equation}
P_{\kappa}(l)=\frac{9H_0^2\Omega_m^2}{4c^2}\int_{\chi_1}^{\chi_2}\mathrm{d}\chi\left(\frac{\chi_s-\chi}{\chi_s a(\chi)}\right)^2 P_{\rm nl}(l/\chi,\chi)\; ,
\end{equation}
where we have assumed a fixed comoving distance of the sources $\chi_s$. For calculating $C^{\rm LSS}$ around our cluster haloes, we set the source redshift to a fixed $z_s=1$ and limit the interval $[\chi_1,\chi_2]$ to the range of $\pm200h^{-1}$ comoving Mpc included in the boxes around the cluster haloes (cf. Section~\ref{sec:sim}). For the non-linear matter power spectrum $P_{\rm nl}$ we use the model of \citet{2003MNRAS.341.1311S} with the \citet{1998ApJ...496..605E} transfer function including baryonic effects. We perform the integral using a customized version of \textsc{nicaea} \citep{2009A&A...497..677K}. The result is shown in the lower left panel of Fig.~\ref{fig:sketch}.

\subsubsection{Concentration}
\label{sec:conc}
The covariance due to variations in concentration parameter at fixed mass is calculated as
\begin{eqnarray}
C^{\rm conc}_{ij}&=&\int \mathrm{d}P(c) \kappa_i\kappa_j \nonumber \\ &-& \left[ \int \mathrm{d}P(c) \kappa_i \right]\times\left[ \int \mathrm{d}P(c) \kappa_j\right] \; ,
\end{eqnarray}
where $\kappa_i$ is taken to be the convergence of a halo of concentration $c$ in annulus $i$. For the probability density $\mathrm{d}P(c)$, we assume a log-normal scatter of $\sigma_{\log_{10} c_{200m}}=0.18$ \citep{2001MNRAS.321..559B} about the mean concentration of \citet{2008MNRAS.390L..64D}. The result is shown in Fig.~\ref{fig:sketch}, lower second-to-left panel. As can be seen, it is subdominant at the radii considered here relative to other sources of intrinsic profile variation, at least for clusters of moderate mass, which illustrates the difficulty of measuring individual cluster concentrations with weak lensing.

\subsubsection{Correlated large-scale structure}
\label{sec:corr}

Clusters of galaxies are likely to form in overdense regions, where the abundance of additional (correlated) haloes is also higher than average. The mean effect of this on projected density is the two-halo term of Section~\ref{sec:twohalo}. However, the stochastic variation of the number of correlated haloes around a cluster also contributes to the covariance of $\bm{\kappa}$. 

We calculate this effect in a halo model, in analogy to \citet[][their Appendix A2]{2011MNRAS.416.1392G}. The idea is to split the set of all possible haloes into subsets that are alike in the sense that they cause a similar shear signal, e.g. ones that are of similar mass and projected distance from the cluster. We then apply Poissonian statistics to the number of haloes that is actually present from each subset.

Let $\mathbb{H}=\lbrace\bm{h}\rbrace$ be the set of tuples $\bm{h}$ that completely characterize all possible halos, e.g. in terms of their coordinates and masses. Consider mutually exclusive and collectively exhaustive subsets $H_k\subset\mathbb{H}$ that are small enough such that $\forall H_k:\;\forall\bm{h}_a,\bm{h}_b\in H_k: \Sigma^i(\bm{h}_a)\approx\Sigma^i(\bm{h}_b)$ is a good approximation for the surface mass density of such a halo averaged over annulus $i$. Then we can write the surface mass density of correlated haloes $\Sigma_{\rm corr}^i$ in annulus $i$ as
\begin{equation}
\Sigma_{\rm corr}^i = \sum_k \lambda_k \Sigma^i(\bm{h}_k) \,
\end{equation}
where $\lambda_k\geq0$ is a random variable that describes the number of correlated haloes $\in H_k$ realized in a manifestation of the cluster. Defining a halo excess probability density $\mathrm{d}P_c$ around a cluster with properties $\bm{h}_{\rm cl}$ such that $\int_{H_k} \mathrm{d}P_c(\bm{h}|\bm{h}_{\rm cl})=\langle\lambda_k\rangle$, this immediately results in the expectation value
\begin{equation}
\langle\Sigma_{\rm corr}^i\rangle = \int \mathrm{d}P_c(\bm{h}|\bm{h}_{\rm cl}) \Sigma^i(\bm{h}) \; .
\label{eqn:correxp}
\end{equation}

For the covariance we make the assumption (discussed below) that the population random variables $\lambda_k$ and $\lambda_l$ of the halos in sets $H_k$ and $H_l$ are Poisson distributed and mutually independent ($\Cov(\lambda_k,\lambda_l)=\delta_{kl}\lambda_k$). Then
\begin{equation}
C^{\rm corr}_{ij}\times\Sigma_{\rm crit}^2 = \int \mathrm{d}P_c(\bm{h}|\bm{h}_{\rm cl})  \Sigma^i(\bm{h}) \Sigma^j(\bm{h}) \; .
\label{eqn:corrint}
\end{equation}

If we characterize secondary haloes only by their mass and projected angular distance from the cluster centre, $\bm{h}=(M,\theta)$, we can write
\begin{eqnarray}
\mathrm{d}P_c(\bm{h}|\bm{h}_{\rm cl}) = b(M_{\rm cl})b(M) \frac{\mathrm{d}N(M,z_{\rm cl})}{\mathrm{d}M\mathrm{d}V} W(\theta, z_{\rm cl}) \, 2\pi\theta\mathrm{d}\theta\mathrm{d}M\; , \nonumber \\
\label{eqn:DeltaSigma}
\end{eqnarray}
Here we have introduced the halo mass function $\frac{\mathrm{d}N(M,z_{\rm cl})}{\mathrm{d}M\mathrm{d}V}$ and used the projected linear excess Lagrangian depth $W$ from eqn.~\ref{eqn:W}, which assumes that pairs of haloes of masses $M_{\rm cl}$ and $M$ cluster according to the linear matter two-point correlation and a mass dependent linear halo bias $b$ as 
\begin{equation}
\xi_{\rm hh}(M_{\rm cl},M, r)=b(M_{\rm cl},z_{\rm cl})b(M,z_{\rm cl})\xi_{\rm lin}(r,z_{\rm cl}) \;.
\end{equation}

We perform the two-dimensional integral of equations~\ref{eqn:corrint} and \ref{eqn:DeltaSigma}, factoring out the cluster halo bias $b(M_{\rm cl},z_{\rm cl})$ for later re-scaling of the covariance matrix to clusters of any mass. For the halo mass function and halo bias we use the models of \citet{2008ApJ...688..709T,2010ApJ...724..878T}. The haloes are modelled as BMO profiles (cf. Section~\ref{sec:onehalo}) with concentration according to \citet{2008MNRAS.390L..64D} and the truncation radius model of Eqn.~\ref{eqn:taum}. We include contributions from haloes of mass $10^{8}h^{-1}\Msol\leq M_{200m}\leq10^{15.5}h^{-1}\Msol$. 

Finally, we note that a mutual (three-point) correlation of different correlated haloes is in fact expected, i.e. the presence of one massive, secondary halo makes the presence of tertiary haloes more likely. This means that the approach presented above yields merely a lower limit of the true covariance, yet this can be approximately compensated by an empirical re-scaling of the covariance matrix with $c^{\rm corr}$. The resulting covariance matrix is shown in the lower second-to-right panel of Fig.~\ref{fig:sketch}.

\subsubsection*{Non-linearly correlated subhaloes}

The model for correlated large-scale structure in Section~\ref{sec:corr} assumed a linear correlation of secondary haloes at all radii. A more realistic approach might be to make this assumption only for haloes outside the virial radius and add a distinct population of subhaloes. We test this model as follows.

For the subhalo abundance inside the virial radius \citep[][their eqn.~6]{1998ApJ...495...80B} we use the subhalo mass function of \citet{2014arXiv1403.6827J}. The surface density of haloes is distributed according to an NFW profile with concentration $c_{200m}=3.9$, adapted from the measurement of $c_{200c}=2.6$ by \citet{2012MNRAS.423..104B} for \citet{2008MNRAS.390L..64D} halo concentration, $M_{200m}=2\times10^{14} h^{-1} \Msol$ and the snapshot redshift. These assumptions yield $\mathrm{d}P_c$ for the subhalo case.

We assume a truncation of subhaloes at their $r_{200m}$ radius, i.e. $\tau_{200m}=c_{200m}(M_{\rm sub})$, motivated by halo stripping. For the projected density profile of the individual subhalo, we ensure mass compensation by subtracting as much matter according to the mean subhalo density profile (eqn.~\ref{eqn:correxp}) as is contained in the individual halo. Otherwise we apply the prescription of Section~\ref{sec:corr} to determine the subhalo shot noise covariance matrix $C^{\rm sub}$.  In this we are ignoring, as before for the linear correlation, both a potential dependence of the subhalo number density profile on the parent halo concentration and ellipticity (which might cause correlations between $C^{\rm sub}$ and the respective other components of the model) and mutual correlation of subhaloes.

The covariance for linearly correlated haloes in this model is calculated as in eqn.~\ref{eqn:corrint}, yet using a virial sphere excised version of $W(\theta)$ to calculate $\mathrm{d}P_c(\bm{h}|\bm{h}_{\rm cl})$ in eqn.~\ref{eqn:DeltaSigma}. Note that $C^{\rm corr}$ in this model explicitly depends on the halo mass rather than being a constant template that is re-scaled by the central halo bias.

We find that the predicted subhalo covariance is a significant contribution to the intrinsic covariance of the halo profile at the high mass end of our sample. The contribution is dominated by massive subhaloes that are sufficiently resolved by our simulations. When we fit a linear dependence of a re-scaling parameter $c^{\rm sub}=c^{\rm sub}_0 + c^{\rm sub}_1(\nu-\nu^{\rm sub}_0)$ together with $c^{\rm corr}_{0/1}$ and $c^{\rm ell}_{0/1}$ as described in Section~\ref{sec:covfit} below, we find $c^{\rm sub}_0=0.8\pm0.8$ and $c^{\rm sub}_1=-2\pm1.5$. Fixing $c^{\rm sub}_1=0$ and $c^{\rm corr}_1=0$ yields a best-fit $c^{\rm sub}_0=0$. The contribution of subhaloes is particularly degenerate with the halo asphericity covariance $C^{\rm ell}$ (see Section~\ref{sec:ell}). We conclude that a larger sample of massive haloes would be required to determine whether the addition of $C^{\rm sub}$ improves the model and therefore do not include it in the following analyses.

\subsubsection{Halo asphericity}
\label{sec:ell}
Dark matter haloes are known to be triaxial in general. The axis ratios and their orientation along the LOS can change the projected $\kappa$ profile significantly. It is difficult, however, to model the effect in its full generality. 

We therefore make the following simplified model for the covariance due to halo asphericity. In accordance with the dominant feature of halo shapes in nature, we assume them to have a prolate shape with minor-to-major axis ratio $0< q=b/a=c/a\leq 1$. We define a coordinate system with ellipsoidal radius $r_e$ such that
\begin{equation}
 r_e^2=\bm{x}^T\left(\begin{array}{ccc}q^{-2/3}&0&0\\0&q^{-2/3}&0\\0&0&q^{4/3}\end{array}\right)\bm{x} \; ,
\end{equation}
where $\bm{x}^T=(x,y,z)$ is a Cartesian coordinate system centred on and aligned with $z$ along the major axis of the halo. For the spherical coordinate system $(R,\theta,\phi)$ with $\theta=0$ along the major axis, 
\begin{equation}
r_e(R,\theta)/R=\sqrt{q^{-2/3}\sin^2\theta+q^{4/3}\cos^2\theta} \; .
\end{equation}

We find the three-dimensional density of a prolate NFW halo by evaluating equation~\ref{eqn:nfw} at $r=r_e$. By virtue of the unit determinant, the volume of ellipsoidal shells is $V(r_e,r_e+\mathrm{d}r_e)=4\pi r_e^2 \mathrm{d}r_e$. The usual normalization of $\rho_0$ therefore means that the density integrated out to an ellipsoidal radius $r_e=r_{200m}$ matches $M_{200m}$. However, we need to re-scale $\rho_0$ numerically for the mass inside a \emph{sphere} of radius $r_{200m}$ to match $M_{200m}$ (cf. \citealt{2002ApJ...574..538J,2003ApJ...599....7O,2007MNRAS.380..149C} for definitions based on ellipsoidal overdensity and \citealt{2014MNRAS.443.1713D} for a spherical overdensity approach to ellipsoidal haloes). 

Integration of $\rho$ along the LOS yields the surface mass density for any combination of $M_{200m}$, $c_{200m}$, $q$ and orientation angle $\alpha$, where $\alpha=0$ puts the major axis along the LOS. We take care to use an approximation to the analytical result for the integrated density inside the innermost region, $r/r_s<10^{-3}$, where the diverging density leads to numerical instability. Here we can approximate $\theta\approx\alpha$ for most of the matter along the LOS to find the mean surface density inside a small projected radius $\delta v=\delta r/r_s$ as (cf. \citealt{1996AuA...313..697B,2000ApJ...534...34W})
\begin{eqnarray}
\langle\Sigma\rangle (<\delta v) = \rho_0 r_s \nonumber \\ \times \left((1-e^2)^{-1/3}\sin^2\alpha+(1-e^2)^{2/3}\cos^2\alpha\right)^{-1/2} \nonumber \\ \times \frac{4}{\delta v^2}\left(\frac{2}{\sqrt{1-\delta v^2}}\rm{arctanh}\left(\sqrt{\frac{1-\delta v}{1+\delta v}}\right)+\ln\left(\frac{\delta v}{2}\right)\right) \; .
\end{eqnarray}

The covariance matrix is integrated as
\begin{eqnarray}
C^{\rm ell}_{ij}&=&\int \mathrm{d}P(q,\cos\alpha) \kappa_i\kappa_j \nonumber \\ &-& \left[ \int \mathrm{d}P(q,\cos\alpha) \kappa_i \right]\times\left[ \int \mathrm{d}P(q,\cos\alpha) \kappa_j\right] \; ,
\end{eqnarray}
where $\kappa_i$ is the mean surface density of a halo of axis ratio $q$ and orientation angle $\alpha$ in annulus $i$. We assume isotropic orientation, i.e. a uniform distribution of $\cos\alpha\in[0,1]$, and a truncated Gaussian distribution of $q$
\begin{equation}
P(q)\propto\left\lbrace \begin{array}{ll}\mathcal{N}(\mu=0.6,\sigma=0.12), & 0.1\leq q\leq 1\\0, & q<0.1 \vee q>1\end{array}\right. \; ,
\end{equation}
approximating the distribution of $q$ for haloes of mass $\approx10^{14}\Msol$ as measured in simulations by \citet[][cf. their Section~4.3]{2007MNRAS.376..215B}. The resulting covariance matrix, re-scaled empirically by $c^{\rm conc}$, is shown in the lower right panel of Fig.~\ref{fig:sketch}. We see that the influence of halo ellipticity is limited to small radii compared to correlated and uncorrelated structures along the LOS, yet quite significant in that regime.

\subsection{Covariance estimation}

\label{sec:ensemble}

Given a model for $\bm{\kappa}(M_{200m})$ that we can subtract from the observed $\bm{K}$, any simulated cluster of mass $M_{200m}$ yields an estimate for $C$ as
\begin{equation}
\Cov(E_i,E_j)=\langle E_i E_j\rangle \; .
\label{eqn:covdef}
\end{equation}
While this approach heavily relies on the accuracy of $\bm{\kappa}$, it is also possible to estimate the covariance matrix $C(M_{200m})$ from an ensemble of $N$ simulated clusters of fixed mass $M_{200m}$ without assuming a model for the mean profile. This is done by applying the ensemble covariance estimator
\begin{equation}
\Cov(E_i,E_j)=\frac{N}{N-1}\langle \hat{E}_i \hat{E}_j\rangle \; ,
\label{eqn:covemp}
\end{equation}
where $\bm{\hat{E}}$ is the residual with respect to the ensemble mean profile. The variance of the estimator of eqn.~\ref{eqn:covemp} is larger than the one of eqn.~\ref{eqn:covdef} by a factor $N/(N-1)$. We decide to bin our clusters in mass in subsamples of $N=24$, such that the loss of information due to not assuming a model for the true mean profile is negligible. However, especially at the massive end where the number of clusters in our simulations is small, this would introduce an additional variance of profiles due to the systematic change of mass within each subsample. We therefore define
\begin{equation}
\hat{E}_i^k=\sqrt{\frac{N}{N-1}}\left(K_i^k-N^{-1}\sum_{l=1}^N K_i^l\times\frac{\kappa_i^k}{\kappa_i^l}\right),
\label{eqn:E}
\end{equation}
where $K_i^j$ and $\kappa_i^j$ are the actual and model convergence of cluster $j$ in radial bin $i$. Here, we have re-scaled each of the other cluster profiles to the expected value at the mass of cluster $k$ by means of the model and amplified the deviations from the mean by $\sqrt{N/(N-1)}$ to correct for the bias of the maximum-likelihood ensemble variance, such that $\Cov(\hat{E}_i,\hat{E}_j)=\Cov(E_i,E_j)$.

\section{Determination of covariance model parameters}

\label{sec:covfit}

The multivariate Gaussian of equation~\ref{eqn:P} corresponds to a log-likelihood
\begin{eqnarray}
-2\ln \mathcal{L}&=&\ln\det C(M)\nonumber \\ &+&(\bm{\kappa}(M)-\bm{K})^{\rm T}C^{-1}(M)(\bm{\kappa}(M)-\bm{K})\nonumber \\ &+&\mathrm{const} \; .
\label{eqn:likelihood}
\end{eqnarray}
This could be interpreted as a likelihood of mass or of the parameters of both the mean profile model $\bm{\kappa}(M)$ and the covariance model $C(M)$. For the latter, the first term on the right-hand side serves as a regularization that prevents run-off of the (co)variance estimate to infinity. In our case, as discussed above, we use a non-parametric profile model to replace $\bm{\kappa}(M)-\bm{K}$ by the $\bm{\hat{E}}$ of equation~\ref{eqn:E} and maximize the likelihood to constrain the parameters of the covariance model.

Equation~\ref{eqn:likelihood} has the disadvantage that the precision matrix $C^{-1}(M)$ and $\det C(M)$ become numerically unstable due to the large conditional number and strong covariance between neighbouring radial bins. Related to this, the uncertainty of empirically estimated off-diagonal components of the covariance matrix is large \citep[e.g.][their eqn.~18]{2013MNRAS.432.1928T}. We therefore decide to use equation~\ref{eqn:likelihood}, however with a diagonal model covariance matrix, i.e. one where all off-diagonal components are set to zero.  We verify, using a toy model, that this yields an unbiased maximum-likelihood estimate of the covariance model parameters. 

For the covariance model, we assume equation~\ref{eqn:Cmodel}. The contribution of concentration becomes important only at large mass and small radii, and rather than (poorly) constraining it from the data we decide to adopt the \citet{2001MNRAS.321..559B} log-normal concentration scatter, i.e. we set $c^{\rm conc}=1$. 

As a baseline for $c^{\rm corr}(\nu)$ and $c^{\rm ell}(\nu)$ we use the mass-independent model
\begin{eqnarray}
c^{\rm corr}(\nu)&=&c^{\rm corr}=\mathrm{const} \nonumber \\
c^{\rm ell}(\nu)&=&c^{\rm ell}=\mathrm{const} \; .
\label{eqn:constmodel}
\end{eqnarray}
As a test for mass dependence we also run our analysis with a model that allows for linear evolution in $\nu$,
\begin{eqnarray}
c^{\rm corr}(\nu)&=&c^{\rm corr}_0+(\nu-\nu^{\rm corr}_0)c^{\rm corr}_1 \nonumber \\
c^{\rm ell}(\nu)&=&c^{\rm ell}_0+(\nu-\nu^{\rm ell}_0)c^{\rm ell}_1 \; ,
\label{eqn:linmodel}
\end{eqnarray}
where we set the pivot points $\nu^{\rm corr}_0=2.5$ and $\nu^{\rm ell}_0=2.8$ to make errors on $c^{\star}_0$ and $c^{\star}_1$ approximately uncorrelated in the bootstrap runs.

\subsection{Results}

\label{sec:results}

The best-fitting values for the parameters of eqn.~\ref{eqn:constmodel} with errors estimated from bootstrapping are $c^{\rm corr}=5.0\pm0.3$ and $c^{\rm ell}=3.7\pm0.3$. Allowing for a linear dependence as in eqn.~\ref{eqn:linmodel} these values are unchanged at best fit and we get no indication (but relatively poor constraints) for a $\nu$ dependence with $c^{\rm corr}_1=-0.04\pm0.88$ and $c^{\rm cell}_1=-0.05\pm0.79$. For the remainder of the analysis, we therefore adopt $\nu$ independent $c^{\rm corr}=5.0$ and $c^{\rm ell}=3.7$.

\begin{figure*}
\includegraphics[width=0.48\textwidth]{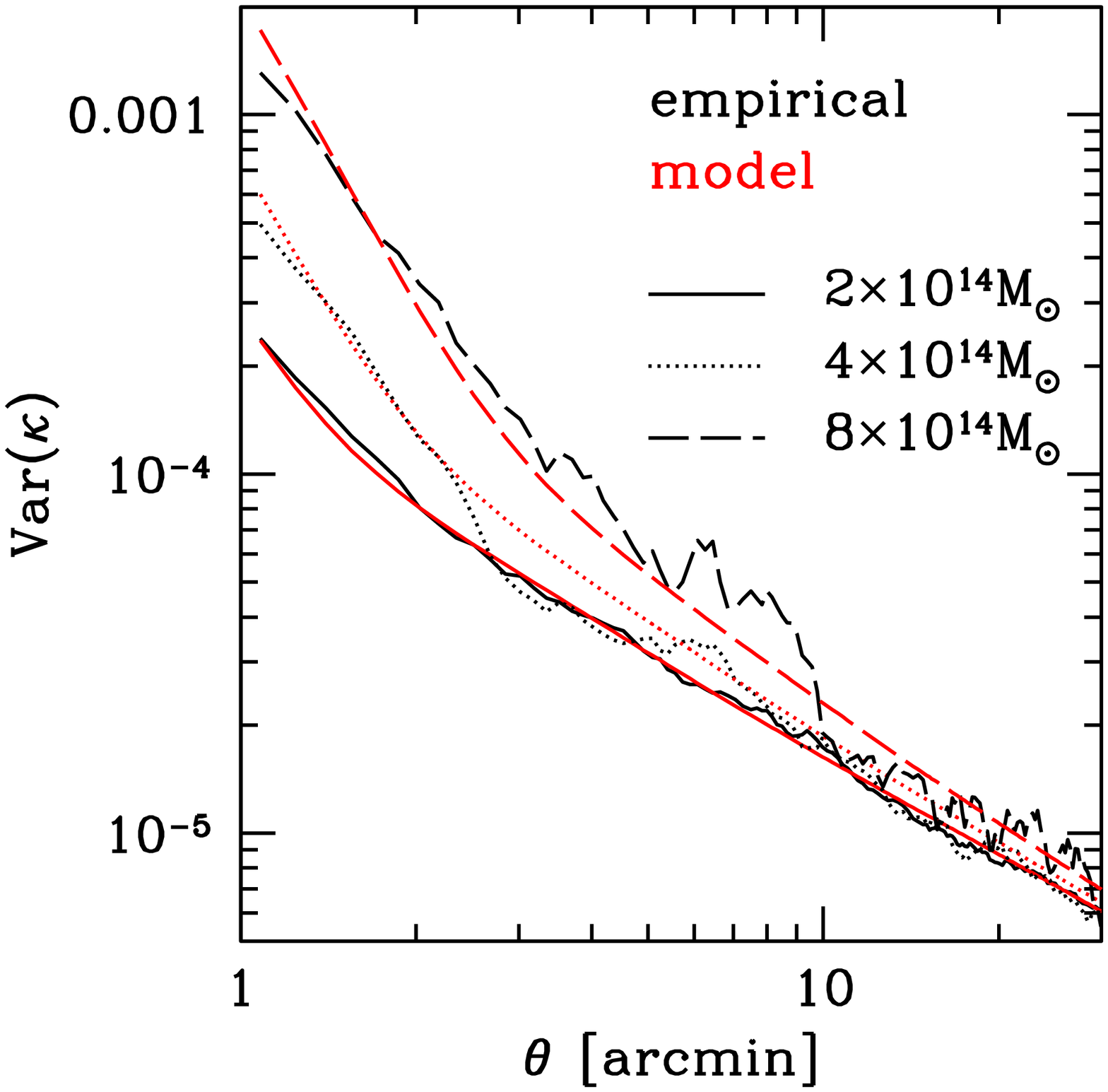}
\includegraphics[width=0.48\textwidth]{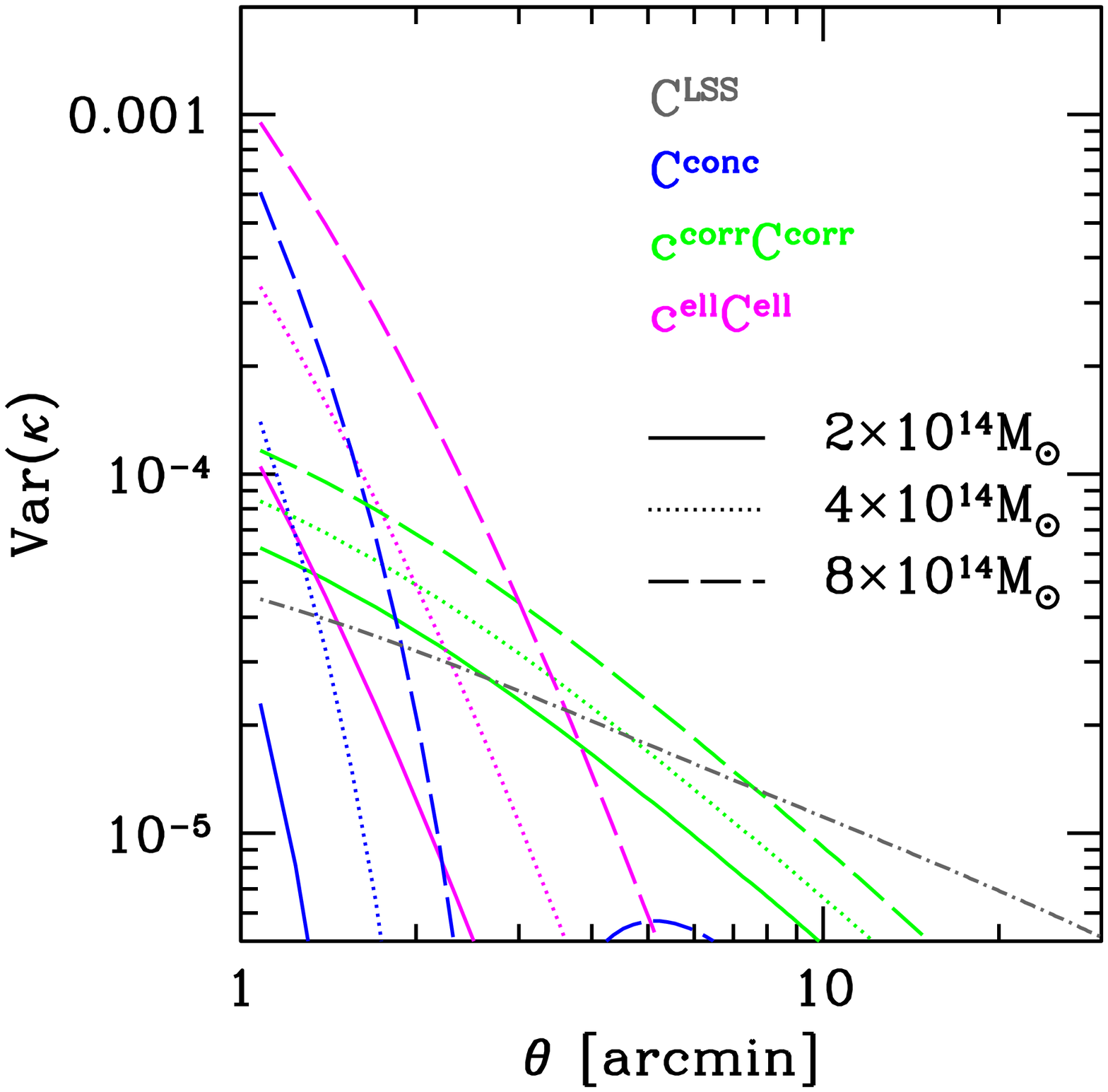}
\caption{Empirical covariance (black, left panel) and covariance model (red, left panel) for clusters of mass $M_{200m}=1.8\ldots2.2\times10^{14}h^{-1}\Msol$ (solid lines), $M_{200m}=3.5\ldots4.5\times10^{14}h^{-1}\Msol$ (dotted) and $M_{200m}=7\ldots9\times10^{14}h^{-1}\Msol$ (dashed lines). Right panel shows components of the model with $c^{\rm corr}=5.0$ and $c^{\rm ell}=3.7$ for all three cases (uncorrelated LSS inside the snapshot box shown as grey, dotted-dashed line).}
\label{fig:var}
\end{figure*}

Figure~\ref{fig:sketch} shows the full covariance model with these parameters for $M_{200m}=2\times10^{14}h^{-1}\Msol$. Figure~\ref{fig:var} shows model and data variance for a range of masses. The model reproduces observed intrinsic variations well over a wide range of mass. We note, however, the large uncertainty in data variance due to our limited sample size.

\subsection{Redshift dependence}

In order to test for a potential redshift dependence of our model parameters, we repeat the analysis with a second snapshot at $z=0.499$, using 788 haloes at $M_{200m}\geq2\times10^{14}h^{-1}\Msol$. 

The best-fitting truncation radii are consistent with eqn.~\ref{eqn:taum}. We fit the parameters of eqn.~\ref{eqn:constmodel} as in Section~\ref{sec:covfit} and determine errors using bootstrapping. Empirical covariances are well reproduced by a model with $c^{\rm corr}=4.2\pm1.1$ and $c^{\rm ell}=3.8\pm0.9$. Due to the much smaller sample size and lower contrast of the haloes relative to LSS noise, the parameters are significantly less well constrained as in the $z=0.245$ snapshot, however, consistent within the uncertainties.

We conclude that at the present level of statistical certainty the best-fit values of Section~\ref{sec:results} can be used over the most relevant redshift range for cluster lensing of $z\approx0.2\ldots0.5$.

\section{Effect on weak lensing cluster surveys}

\label{sec:effect}

In this section, we estimate the influence of intrinsic covariance of density profiles at fixed mass 
\begin{itemize}
 \item on mass measurements of individual clusters (Section~\ref{sec:mass})
 \item and on the determination of parameters of the mass-observable relation (MOR) in a lensing follow-up of a sample of clusters (Section~\ref{sec:mor}).
\end{itemize}

In our analyses we fix the lens redshift to $z_l=0.24533$, the primary redshift of our simulated cluster profiles. We consider shear surveys of different depths, parametrized by the surface density $\mathfrak{n}$ of source galaxies, which we assume to lie at a fixed $z_s=1$. The three settings chosen roughly correspond to current large ground based surveys ($\mathfrak{n}=10$~arcmin$^{-2}$), the best available ground-based data ($\mathfrak{n}=50$~arcmin$^{-2}$) and space-based data ($\mathfrak{n}=100$~arcmin$^{-2}$).

The full covariance of equation~\ref{eqn:Cmodel} contains, apart from uncorrelated and intrinsic variations of surface mass density, the measurement uncertainty of the survey $C^{\rm obs}$. For a shear survey, assume that we can measure the tangential gravitational shear $\bm{\gamma}$ in each annulus $i$ with variance
\begin{equation}
 \sigma^2_{\gamma,i}=\frac{\sigma^2_{\epsilon}}{\mathfrak{n}A_i} \; ,
\end{equation}
where $\sigma_{\epsilon}\approx0.3$ is the shape noise (including intrinsic shape dispersion and measurement noise), $\mathfrak{n}$ is the background source density and $A_i$ the area of annulus $i$. The $n$-dimensional covariance matrix of $\bm{\gamma}$ (ignoring intrinsic alignment and shear systematics) is diagonal with $C^{\mathrm{obs},\gamma}_{ii}=\sigma^2_{\gamma,i}$.

In the limit that annuli are thin, $\bm{\gamma}$ and $\bm{\kappa}$ are connected by a linear equation $\bm{\gamma}=G\bm{\kappa}$, where the $n\times n$ matrix $G$ is defined such that
\begin{equation}
 \gamma_i=\langle\kappa\rangle_{<i}-\kappa_i=\sum_{j=0}^{i-1}\kappa_j A_j / (\pi \theta_{i,\mathrm{min}}^2)-\kappa_i \; .
\end{equation}
Our goal is to calculate the observational covariance $C^{\mathrm{obs}}$ of $\kappa$, which can be written as
\begin{equation}
 C^{\mathrm{obs}}=G^{-1} C^{\mathrm{obs},\gamma} (G^{-1})^T \; .
\end{equation}

The above derivation has omitted the technical step of breaking the mass-sheet degeneracy, which is necessary for $G$ to be of full rank. To this end, we increase the dimensionality by 1 and let the newly introduced entry $\gamma_{n+1}$ represent the convergence in the outermost radius, and the new $\kappa_{n+1}$ be the mean convergence inside the innermost annulus. The measurement error of $\gamma_{n+1}$ is defined to be the LSS variance of $\kappa$ on the outermost angular scale (this is essentially what happens in the common assumption of $\kappa=0$ on large scales). The mean convergence inside the innermost annulus is connected to $\bm{\gamma}$ by the linear equation above. We later exclude the added component from our analysis again, i.e. we consider only the sub-matrix $i,j\in[1,n]$ (similar in effect to \citealt{2010MNRAS.405.2078M}).

Apart from the \textit{full model covariance} $C(M)$ of equation~\ref{eqn:Cmodel}, we also define the \textit{covariance without intrinsic variations}
\begin{equation}
C^{\rm noint}=C^{\rm obs}+C^{\rm LSS} \; .
\label{eqn:Cnoint}
\end{equation}

For a model independent representation, we calculate the ensemble covariance of the re-scaled profiles (cf. eqn.~\ref{eqn:E}) as $\hat{C}^{\rm int}(M)$. When calculating $\hat{C}^{\rm int}(M)$, we use 100 clusters that are nearest neighbours in a mass-ordered list of our simulated haloes to the target mass $M$ and subtract the uncorrelated LSS covariance inside the simulation box. The \textit{empirically estimated covariance} is
\begin{equation}
C^{\rm emp}(M)=C^{\rm obs}+C^{\rm LSS}+\hat{C}^{\rm int}(M) \; .
\label{eqn:Cemp}
\end{equation}

\subsection{Mass confidence intervals}
\label{sec:mass}

We determine the effect of intrinsic variations of the density profile on the validity of confidence intervals of weak lensing mass measurements. In this procedure, we assume the empirically estimated density profile covariance from our ensemble of simulated haloes to represent the true variability. We test for the effects of either taking into account intrinsic variations with the model proposed in this work or ignoring them. We design this test as follows.

At given mass $M$, we define and determine the full empirically estimated covariance $C^{\rm emp}(M)$ as described in eqn.~\ref{eqn:Cemp}. In a Monte Carlo simulation, we then generate profiles by adding a multivariate Gaussian random vector according to $C^{\rm emp}(M)$ to the model profile $\bm{\kappa}(M)$ (eqn.~\ref{eqn:profilemodel}) of mass $M$.\footnote{As a test for the validity of considering only multivariate Gaussian variations over the mean profile, we perform another run where instead of a synthetic variation we use 100 nearest neighbors in mass around $M$, re-scale their profile to $M$ using the model profiles of eqn.~\ref{eqn:profilemodel} and subtract their mean to generate 100 realistic random variations. We then add these (and multivariate Gaussian random vectors to simulate observational and uncorrelated LSS covariance) to the model profile at $M$ and run the likelihood as described below. Results in this approach do not differ significantly from the synthetic multivariate Gaussian ones presented here.} 

Consider one such realization $\bm{K}$. The likelihood in eqn.~\ref{eqn:likelihood}, now interpreted as a function of mass $\mathcal{L}(M'|\bm{K})$, can then be run over a range of model profiles $\bm{\kappa}(M')$ with different mass $M'$. For $C(M')$ and $C^{-1}(M')$ we use
\begin{itemize}
 \item either the covariance without intrinsic variation $C^{\rm noint}$ of equation \ref{eqn:Cnoint} (colour coded red in the following figures)
 \item or the covariance of equation \ref{eqn:Cmodel}, including our parametric model for intrinsic variations (colour coded blue). 
\end{itemize}
Note that the latter option requires to include the $\ln \det C$ term of the likelihood, since covariance is now a function of mass.

We determine 68\% (and 90\%) confidence intervals using mass limits where the likelihood is $2\Delta\ln\mathcal{L}=1$ (and 2.7) worse than at best fit. We repeat the procedure for 10,000 realizations at every fixed mass and determine the empirical coverage, e.g. the fraction of cases in which the true mass lies inside these confidence intervals.

\begin{figure}
\centering
\includegraphics[width=0.36\textwidth]{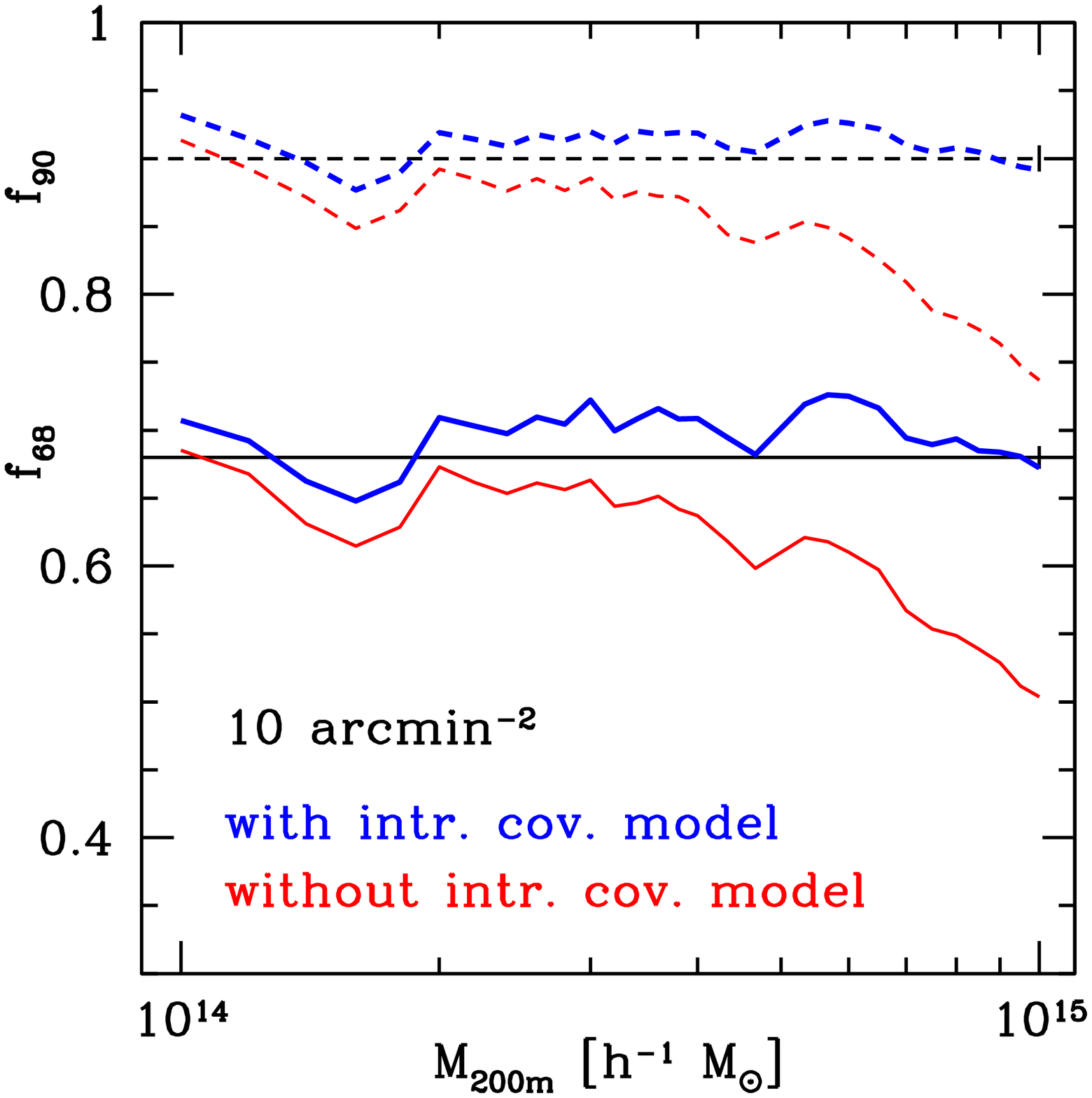}
\includegraphics[width=0.36\textwidth]{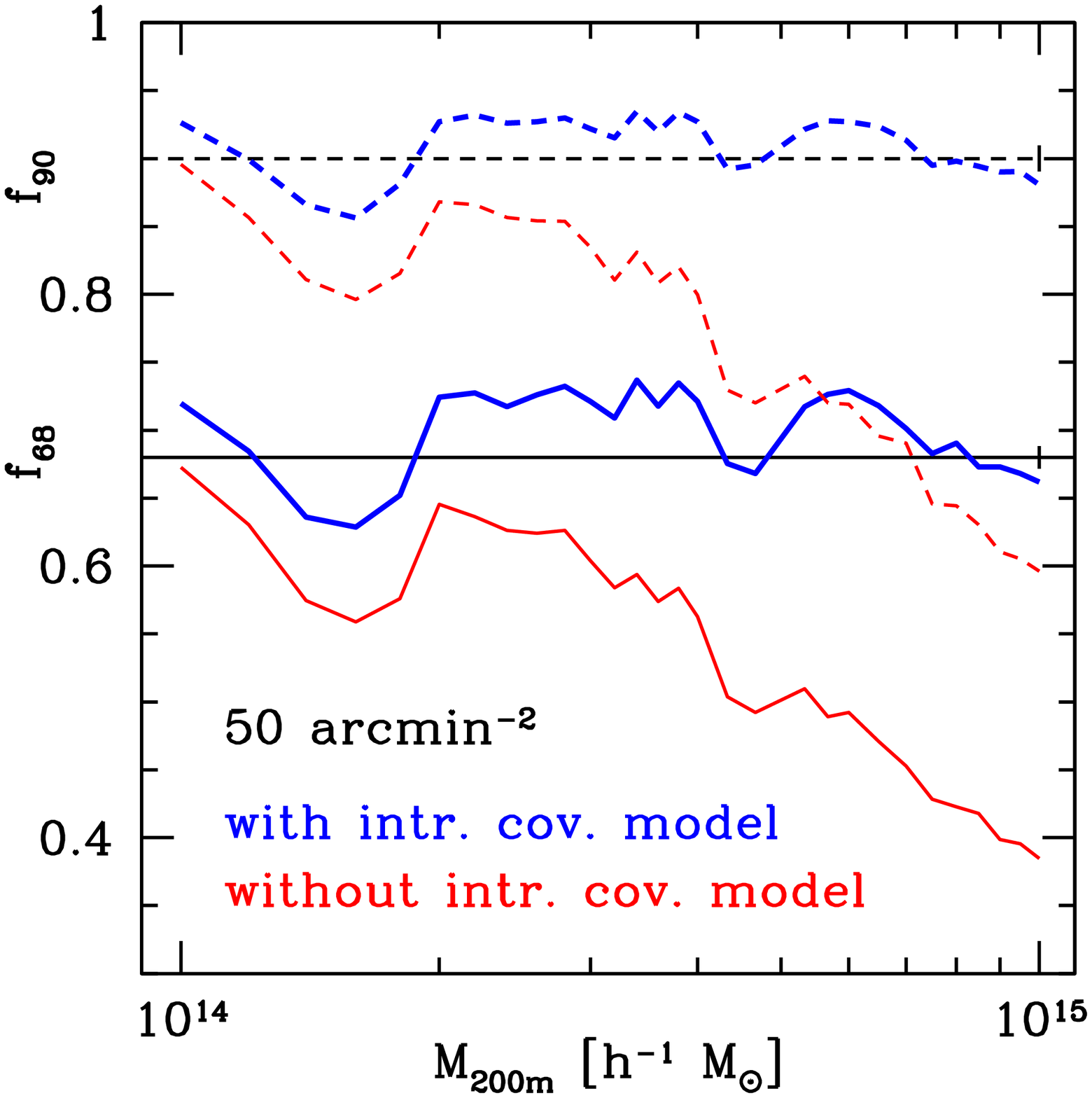}
\includegraphics[width=0.36\textwidth]{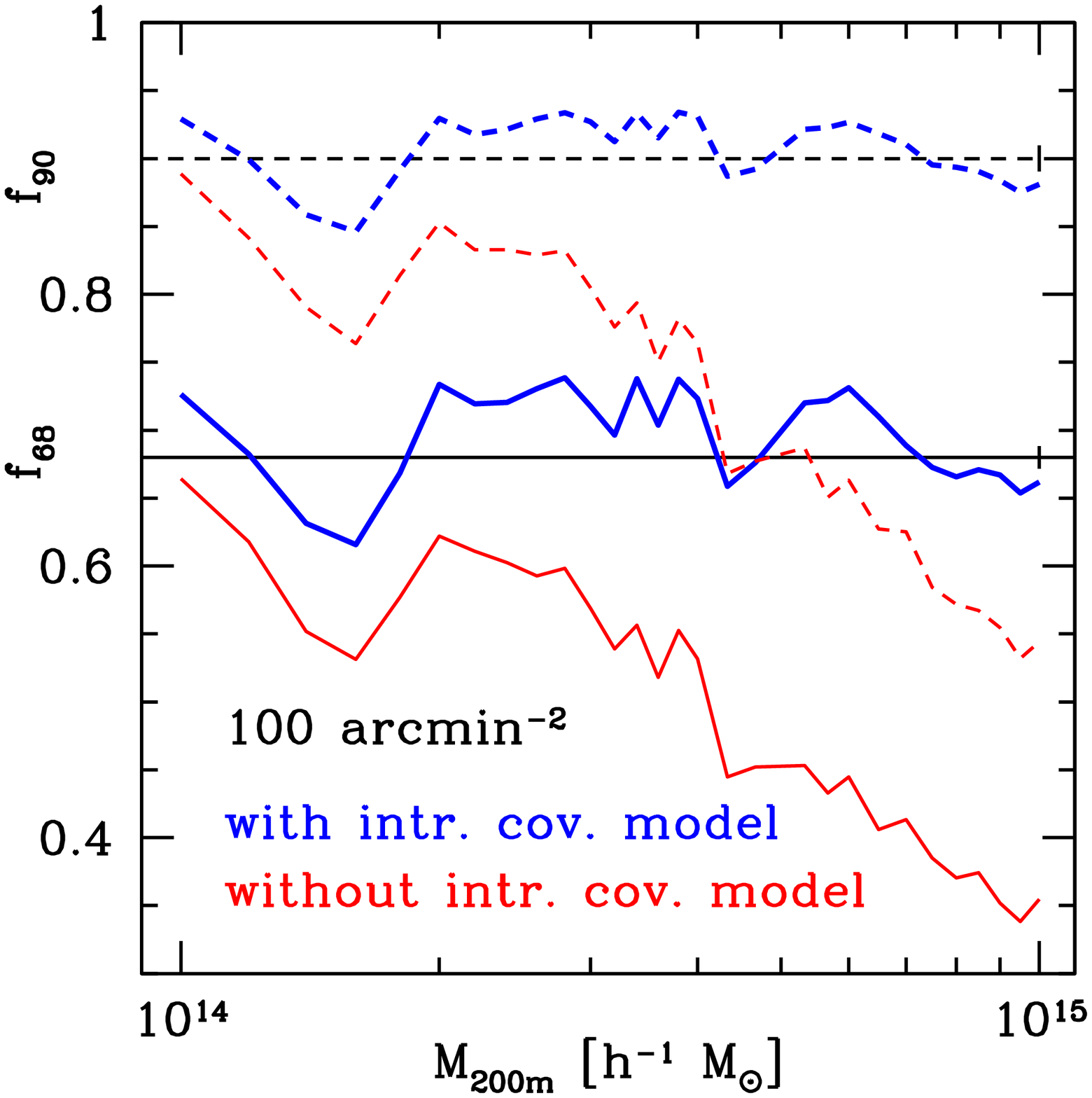}
\caption{Fraction of weak lensing measurements of cluster mass where the true mass is inside the 68\% ($f_{68}$, solid lines) and 90\% ($f_{90}$, dashed lines) confidence interval (black lines at target levels for reference). In red, thin lines we show the common case where only measurement noise and uncorrelated LSS are considered for the covariance matrix (eqn.~\ref{eqn:Cnoint}). Blue, thick lines show results that include our model for the intrinsic covariance (eqn.~\ref{eqn:Cmodel}). Panels correspond to typical large ground based-surveys (top), best available ground based data (centre) and space-based analyses (bottom). As data quality and cluster mass increase, intrinsic variations become a significant component of the uncertainty budget of weak lensing mass measurements and should not be ignored. All plots are for lenses at $z_l=0.24533$.}
\label{fig:massconf}
\end{figure}

Results for three different survey depths are shown in Figure~\ref{fig:massconf}, where we now compare results of the two methods (red and blue lines) to the target levels of 68 and 90\% (black lines). Increasingly with increasing mass, ignoring intrinsic variations leads to an underestimation of errors. Even for present surveys with relatively shallow depth, this manifests in $\approx15$\% excess outliers at the one and two sigma level for massive clusters at around $M_{200m}\approx10^{15} h^{-1} \Msol$. The effect is more severe in deeper data, where the relative importance of intrinsic over observational covariance increases. 

Using our model covariance, confidence intervals are correctly estimated over a wide range of masses and even in deep data. The small number of systems above $M_{200m}>10^{15} h^{-1} \Msol$, however, limits the range of masses where we can evaluate the validity of our model. Note in this context that the scatter in the plots is due to the noise in $\hat{C}^{\rm int}(M)$, as estimated on a limited number of clusters, rather than the number of realizations we run (the effect of the latter is indicated as the small vertical bar on the black lines). 

\subsubsection{Effect of intrinsic covariance on mass uncertainty}

In the previous section, we found that the confidence interval based on a $\bm{\kappa}$ covariance matrix without contributions from intrinsic profile variation is significantly too narrow, especially for more massive clusters and larger depth of the data. Here, we perform a Fisher analysis to determine how well masses can be measured, a question particularly relevant for the design of lensing follow-up programmes.

For a mass dependent covariance matrix, the Fisher information for mass $\mathcal{F}$ reads \citep[e.g.][]{1996ApJ...465...34V,1997ApJ...480...22T}
\begin{equation}
 \mathcal{F}(M)=\left(\frac{\mathrm{d}\bm{\kappa}}{\mathrm{d}M}\right)^T C^{-1} \left(\frac{\mathrm{d}\bm{\kappa}}{\mathrm{d}M}\right) + \frac{1}{2}\tr\left[\left(C^{-1}\frac{\mathrm{d}C}{\mathrm{d}M}\right)^2 \right] \; ,
\label{eqn:fisher}
\end{equation}
where $\bm{\kappa}$, $C$ and their derivatives are taken to be evaluated at mass $M$.

We calculate the Fisher information for the model without intrinsic variations (eqn.~\ref{eqn:Cnoint}) and our full model. In the first case, the second term of eqn.~\ref{eqn:fisher} drops because of the mass independence of the covariance. Including intrinsic variations, we also evaluate the second term, which however contributes to the Fisher information at most at the per-cent level. The mass uncertainty $\sigma_{M}$ is related to the Fisher information as $\sigma_{M}=\mathcal{F}^{-1/2}$.

\begin{figure}
\centering
\includegraphics[width=0.48\textwidth]{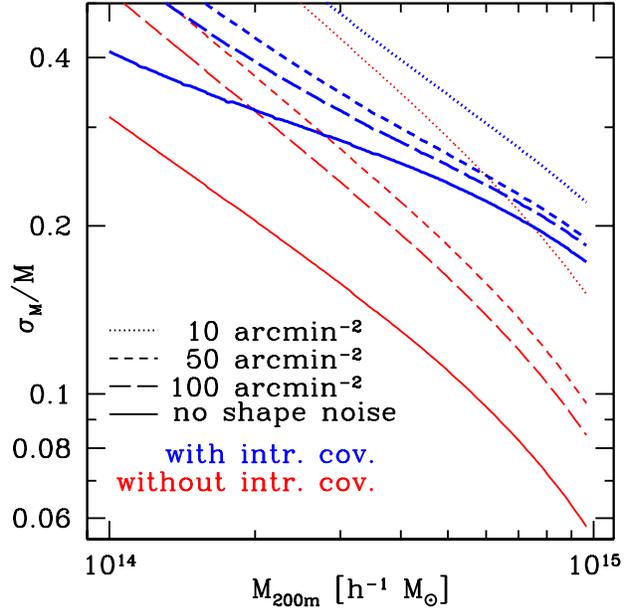}
\caption{Relative uncertainty of weak lensing mass measurements from a Fisher analysis excluding (red) and including (blue) intrinsic covariance according to our model. Not including intrinsic covariance leads to an underestimation of uncertainty, especially at high mass and large depth. Notably, the improvement of uncertainty for massive clusters going from typical ($10$~arcmin$^{-2}$ background galaxies, dotted line) to noiseless measurements (solid line) is only modest when intrinsic variations are considered. Results are for lenses at $z_l=0.24533$.}
\label{fig:fisher}
\end{figure}

Figure~\ref{fig:fisher} shows relative uncertainties in mass $\sigma_{M}/M$ due to a shear surveys of various depth. The increase in uncertainty due to intrinsic profile variation is substantial, even for shallow data with sufficiently massive clusters, but more strongly for deeper observations. Our results are consistent with the findings of \citet{2010arXiv1011.1681B}, who used the same simulations to derive uncertainties empirically rather than with a covariance model.

\subsection{Mass-observable relations}
\label{sec:mor} 

We study the effect of intrinsic variations of projected cluster density profiles on weak lensing follow-up surveys used for determining parameters of a mass-observable relation. Consider an observable $Y$ (which could be the Compton decrement, an X-ray mass proxy or an optical richness) with a power-law mass-observable relation that describes the fiducial value $Y_0(M)$ as a function of mass $M$,
\begin{equation}
\ln Y_0(M)/\hat{Y} = A + B\ln M/\hat{M} \; .
\label{eqn:mor}
\end{equation}
We have introduced here pivot values $\hat{Y}$ and $\hat{M}$ in addition to the power-law slope $B$ and amplitude $A$.

The observed value of $Y$ for any cluster shall include a log-normal intrinsic scatter $\sigma_{\rm int}$ and, for simplicity, also log-normally distributed measurement related uncertainty $\sigma_{\rm obs}$. The combined uncertainty $\sigma=\sqrt{\sigma^2_{\rm int}+\sigma^2_{\rm obs}}$ leads to an observable
\begin{equation}
\ln Y(M) = \ln Y_0(M)+\mathcal{N}(\mu=0,\sigma)\; .
\label{eqn:mor2}
\end{equation}

\subsubsection{Likelihood}

Consider a single cluster with observed convergence profile $\bm{K}$ and measured observable $Y$. We maximize, as a function of MOR parameters, the likelihood
\begin{eqnarray}
P(\bm{K}|Y,A,B,\sigma_{\rm int})= \int P(\bm{K}|M)P(M|Y,A,B,\sigma_{\rm int})\; \mathrm{d}M  \nonumber \\
\propto\int P(\bm{K}|M)P(Y|M,A,B,\sigma_{\rm int})P(M)\; \mathrm{d}M \; .
\end{eqnarray}
For  $P(\bm{K}|M)$ we insert the expression of eqn.~\ref{eqn:P} with $C$ equal to either our full model or the covariance without intrinsic variations. The term for the $Y$ likelihood can be written as
\begin{eqnarray}
P(Y|M,A,B,\sigma_{\rm int})&=&\frac{\exp[-(\ln Y_0(M)-\ln Y)^2/(2\sigma^2)]}{\sqrt{2}\sigma} \nonumber \\ 
&\times&\Theta(Y-Y_{\rm lim})P_{\rm det}^{-1}(M) \; .
\end{eqnarray}
Here we have used the Heaviside step function $\Theta$ to impose the observable limit $Y_{\rm lim}$ and re-normalized by the inverse of the detection probability $P_{\rm det}(M)=\erfc[(\ln Y_{\rm lim}-\ln Y_0(M))/(\sqrt{2}\sigma)]/2$ to compensate for the removed part of the probability distribution \citep[cf.][their eqn.~A10]{2009ApJ...692.1033V} and correct for the (Malmquist) bias due to preferential selection of objects with positive contribution from scatter.

Finally, the normalized mass prior $P(M)$ can be written for an observable limited survey as
\begin{equation}
P(M)\propto \frac{\mathrm{d}n}{\mathrm{d}M}(M)P_{\rm det}(M) \left[\int \mathrm{d}M' \frac{\mathrm{d}n}{\mathrm{d}M}(M')P_{\rm det}(M') \right]^{-1} \; 
\end{equation}
to correct for (Eddington) bias due to the increase of abundance with decreasing mass. Note that $P(M)$ depends on the MOR by means of $P_{\rm det}$ and, by means of the halo mass function $\frac{\mathrm{d}n}{\mathrm{d}M}$, also on cosmology (for which we, however, assume fixed values in the simulations presented here).

\subsubsection{Results}

We simulate samples of clusters $1\leq i\leq n$, where cluster $i$ is characterized by its observable $Y_i$ and an observed convergence profile $\bm{K}_i$. 

For $Y$, we assume a MOR with $B=5/3$ (the self-similar slope for Compton decrement $Y_{\rm SZ}$ and the X-ray equivalent $Y_{\rm X}$) and choose the pivots such that measurement errors on $A$ and $B$ are uncorrelated and $A=0$. We assume an intrinsic scatter $\sigma_{\rm int}=0.1$ and observational uncertainty of $\sigma_{\rm obs}=0.05$. 

The convergence profile $\bm{K}$ is simulated including a multivariate Gaussian deviation from eqn.~\ref{eqn:profilemodel}, according to the full covariance (including uncorrelated LSS, measurement uncertainty of the survey and intrinsic variation as predicted from our model) of eqn.~\ref{eqn:Cmodel}. 

We take the sample to be observable limited, i.e. draw clusters with abundance proportional to the halo mass function, assign observables according to eqns.~\ref{eqn:mor}-\ref{eqn:mor2} and only accept objects where the observable exceeds the survey threshold, $Y>Y_{\rm lim}$. The threshold $Y_{\rm lim}$ is chosen as $Y_{\rm lim}=Y_0(M_{200m}=4\times10^{14}h^{-1}\Msol)$, comparable for instance to ongoing SZ surveys. The number of clusters drawn is taken to be representative of a volume of $V=0.15 h^{-3} $Gpc$^3$ (comoving, at $z=0.24533$), which contains roughly 100 detections in our simulated survey.

\begin{figure*}
\includegraphics[width=\textwidth]{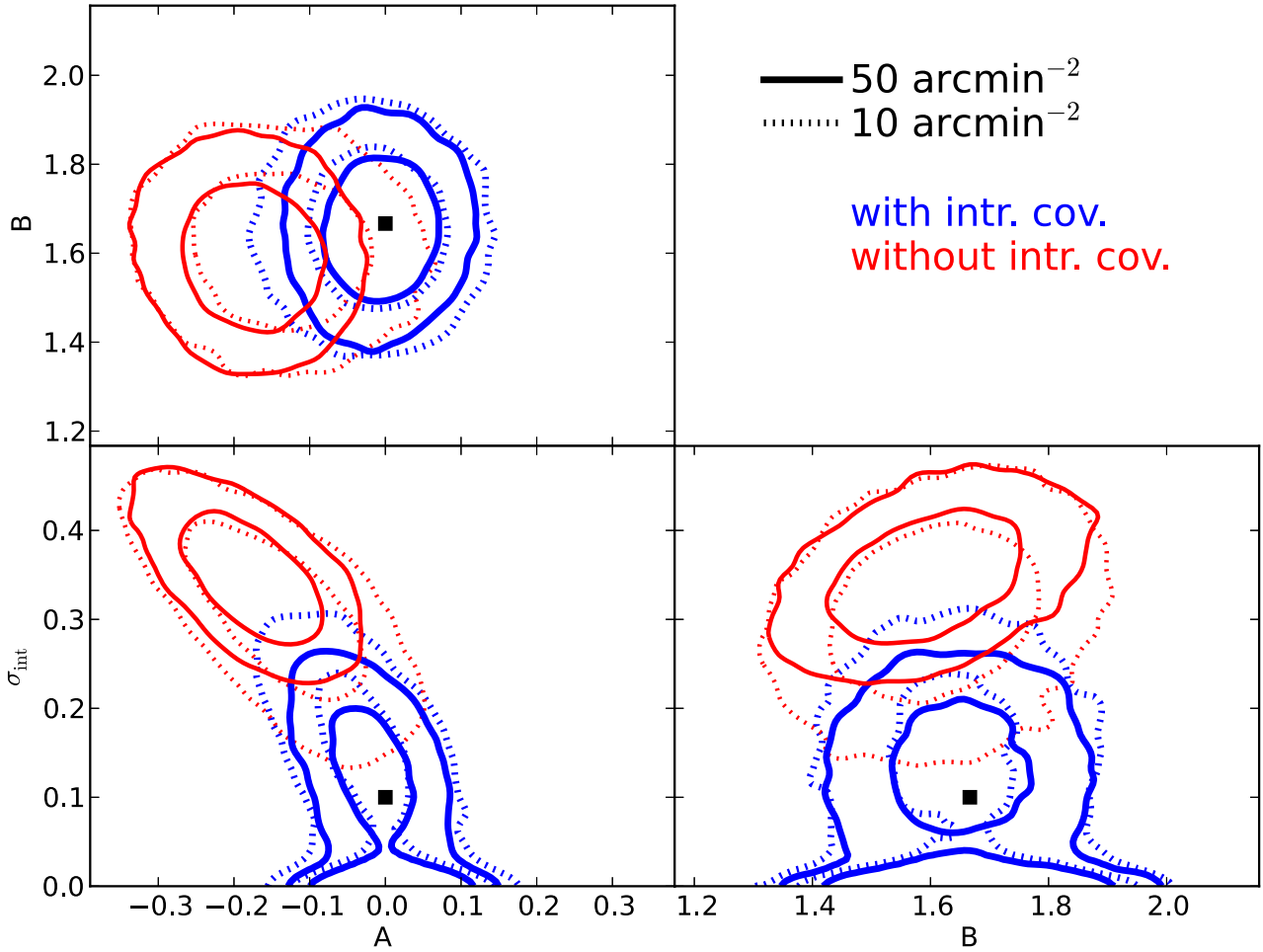}
\caption{Distribution of maximum likelihood MOR parameters (true input indicated by black squares) estimated from a weak lensing follow-up of an observable limited sample. We show results for a shallow (dotted contours, $\mathfrak{n}=10$~arcmin$^{-2}$) and deeper (solid contours, $\mathfrak{n}=50$~arcmin$^{-2}$) lensing survey. Cluster profiles are simulated including intrinsic variation according to our model and the lensing likelihood for cluster mass is estimated including (blue, thick lines) and not including (red, thin lines) the intrinsic component of the covariance matrix in addition to shape noise and uncorrelated LSS. Inner and outer contours enclose 68\% and 95\% of data points, respectively. Results are for lenses at $z_l=0.24533$.}
\label{fig:mor}
\end{figure*}

Figure~\ref{fig:mor} shows the distribution of maximum likelihood estimates of the MOR parameters $A$, $B$ and $\sigma_{\rm int}$, with contours enclosing 68 and 95 per cent of the 15,000 realizations, respectively. We find that the analysis using the full covariance (blue) reproduces the input MOR well. The feature near $\sigma_{\rm int}=0$ is due to the hard prior $\sigma_{\rm int}>0$, which pushes random realization with lower empirical scatter towards this limit. The analysis using the covariance without intrinsic variations (red) significantly overestimates intrinsic scatter. As an additional bias, the overall mass scale $A$ (and, although less strongly, the slope $B$) are systematically underestimated. The move from a shallow (dotted) to deep (solid lines) surveys only moderately improves constraints and makes the bias due to using the covariance without intrinsic variations more apparent.

We note that the biases in $A$ and $B$ are due to a degeneracy between them and the intrinsic scatter. If the latter could be constrained externally with small uncertainty, e.g. by means of realistic simulations or by combining several observables with uncorrelated intrinsic scatter, the bias in the former would be mitigated even without a realistic model for the lensing covariance.

\section{Conclusions}

\label{sec:summary}

We have presented a model for the variation of projected density profiles of clusters of galaxies at fixed mass, constructed as a combination of the effects of variations of halo concentration, ellipticity and orientation, and correlated secondary structures.

The full covariance including our model for intrinsic variations faithfully reproduces confidence intervals in the weak lensing likelihood of cluster mass. We show that when intrinsic variations are ignored, uncertainties in lensing-derived mass are underestimated significantly (cf. Fig.~\ref{fig:massconf}).

Using the full covariance model we have made Fisher predictions for the accuracy of lensing measurements of cluster mass. We have shown that intrinsic variations take away some of the comparative advantage of studying a small sample of the most massive clusters with the deepest possible observations. For a massive cluster ($M_{200m}\approx10^{15}h^{-1}\Msol$ at $z=0.25$) we find an irreducible relative uncertainty in lensing mass of $\approx20$ per-cent due to intrinsic profile variations and uncorrelated LSS along the LOS, three times higher than the uncertainty from uncorrelated LSS alone. Our results agree with the analysis of \citet{2010arXiv1011.1681B}.

With simulations of mock surveys for constraining cluster mass-observable relations with lensing data, we have shown that intrinsic variations significantly bias the derived intrinsic scatter and amplitude if they are not accounted for in the lensing mass likelihood. For a follow-up of a sample of 100 clusters selected by $Y_{SZ}$ above the fiducial value of the observable at $M_{200m}=4\times10^{14}h^{-1}\Msol$ at a redshift $z=0.25$, the bias in the MOR amplitude is $\approx15$ per cent unless tight external constraints on the intrinsic scatter are available. Accounting for the cosmic variance of cluster lensing is therefore necessary for upcoming cluster surveys that target the calibration of MORs for cluster cosmology.

\section*{Acknowledgements}

We thank Gary Bernstein, Eduardo Rozo and Peter Schneider for helpful discussions and Martin Kilbinger for support with the software \textsc{nicaea} used in this work.

This work was supported by SFB-Transregio 33 'The Dark Universe' by the Deutsche Forschungsgemeinschaft (DFG) and the DFG cluster of excellence 'Origin and Structure of the Universe'.

\addcontentsline{toc}{chapter}{Bibliography}
\bibliographystyle{mn2e}
\bibliography{literature}

\label{lastpage}

\end{document}